\documentclass[aps,prb, twocolumn,longbibliography,amsmath,amssymb,nofootinbib,longbibliography,10pt]{revtex4-2}
\usepackage[caption=false]{subfig}
\usepackage{amsfonts,amssymb,amsmath}
\usepackage{graphics}
\usepackage{graphicx}
\usepackage{nicefrac}
\usepackage{cases}
\usepackage{mathrsfs}
\usepackage{enumerate}
\usepackage{mathtools}
\usepackage{textcomp}
\usepackage{verbatim}
\usepackage{bm}          
\usepackage{soul}
\usepackage{cancel}
\usepackage{upgreek}
\usepackage{txfonts}
\usepackage{color}
\usepackage[colorlinks=true, urlcolor=blue, linkcolor=blue]{hyperref}
\usepackage[papersize={8.5in,11in}]{geometry}
\usepackage{esint}
\renewcommand{\vec}[1]{\boldsymbol{#1}}
\newcommand {\be} {\begin{equation}}
\newcommand {\ee} {\end{equation}}
\newcommand {\e} {\varepsilon}

\usepackage{makecell}

\begin{document}

\title{Wigner crystallization in Bernal bilayer graphene}

\author{Sandeep Joy$^{1,2,3,4}$}
\author{Brian Skinner$^{1}$}
\affiliation{$^{1}$Department of Physics, Ohio State University, Columbus, OH 43210, USA}
\affiliation{$^{2}$National High Magnetic Field Laboratory, Tallahassee, Florida 32310, USA}
\affiliation{$^{3}$Department of Physics, Florida State University, Tallahassee, Florida 32306, USA}
\affiliation{$^{4}$FSU Quantum Initiative, Florida State University, Tallahassee, Florida 32306, USA}

\date{\today}
\begin{abstract}

In Bernal bilayer graphene (BBG), a perpendicular displacement field flattens the bottom of the conduction band and thereby facilitates the formation of strongly correlated electron states at low electron density. Here, we focus on the Wigner crystal (WC) state, which appears in a certain regime of sufficiently large displacement field, low electron density, and low temperature. We first consider a model of BBG without trigonal warping, and we show theoretically that Berry curvature leads to a new kind of WC state in which the electrons acquire a spontaneous orbital magnetization when the displacement field exceeds a critical value. We then consider the effects of trigonal warping in BBG, and we show that they lead to an unusual ``doubly re-entrant" behavior of the WC phase as a function of density. The rotational symmetry breaking associated with trigonal warping leads to a nontrivial ``minivalley order" in the WC state, which changes abruptly at a critical value of displacement field. In both cases, we estimate the phase boundary of the WC state in terms of density, displacement field, and temperature. This paper is complementary to our recent work in Ref.~\cite{joy2025chiral}.

\end{abstract}
\maketitle

\section{Introduction}

The confluence of large electronic density of states and nontrivial quantum geometry has led during the past decade to a remarkably rich set of new electronic phenomena. The experimental discovery of Mott insulating physics and superconductivity in twisted bilayer graphene \cite{cao_correlated_2018, Cao_unconventional_2018} has served in many ways as a catalyst for a broader research program that examines two-dimensional electron systems having nontrivial Berry phase physics and tunable dispersion relations. Indeed, studies of twisted bilayer and trilayer graphene have uncovered cascades of phase transitions between different correlated insulating and superconducting phases \cite{Xie_spectroscopic_2019, Serlin_Intrinsic_2020, Stepanov_untying_2020, Zondiner2020cascade, Wong2020cascade, Park_Flavour_2021, choi2021interactiondriven, hao2021electric, Zhou2021Superconductivity, Nuckolls2023quantum, Shen2023Dirac, Kim2023magic, Chen2024strong, Zhou2025gate, Zhang2026angular}. This dramatic success has motivated condensed matter physicists to reconsider untwisted graphene multilayers, in which the band structure can be altered (and made flat near the band edges) by a perpendicular electric field (see, e.g., Refs.~\cite{mccann2006landau, latil2006charge, lu2006influence, ohta2006controlling, ohta2007interlayer, castro2007biased, Zhang2009direct, mak2009observation, koshino2010interlayer, Yankowitz2013local, marchenko2018extremely}). Similarly remarkable results have been uncovered in this setting, including cascades of isospin polarization transitions \cite{Zhou2021half, zhou2022isospin, seiler2022quantum, de2022cascade, lin2023spontaneous, Liu2024spontaneous, guo2025flat}, (unconventional) superconductivity \cite{Zhou2021Superconductivity, zhou2022isospin, zhang_enhanced_2023, tian2023evidence, Li2024tunable, Holleis2025nematicity, Choi2025superconductivity, Han2025signatures, guo2025flat},  quantum anomalous Hall/Chern-insulating states~\cite{Geisenhof2021quantum, Han2023orbital, Han2024correlated, sha2024observation, han2024large,  Choi2025superconductivity, Lu2025extended}, and fractional quantum anomalous Hall/fractional Chern-insulating states~\cite{lu2024fractional, Choi2025superconductivity, Xie2025tunable}.

These recent discoveries have also helped to revive interest in perhaps the oldest theoretically-proposed correlated insulating phase: the Wigner crystal (WC), which was first proposed theoretically in 1934 \cite{Wigner1934On}. Recent experiments are strongly suggestive of the realization of zero-magnetic-field Wigner crystal states in untwisted graphene multilayers \cite{Nam2018afamily, seiler2022quantum, seiler2023interactiondriven, Seiler2025signatures, han2026evidence} (as well as in other two-dimensional materials \cite{hossain2020observation, smolenski_signatures_2021, falson_competing_2022, sung2025electronic, xiang2025imaging}). These experiments have prompted theorists to re-examine basic questions about the nature of the Wigner crystal phase and its quantum freezing/melting transition (e.g., Refs.~\cite{spivak2003phase, spivak2004phases, jamei2005universal, drummond2009phase, Kim2022interstitial, Joy2022Wigner, joy2023upper, kim2024dynamical, huang2024electronic, jain2025elementary, valenti2025critical, esterlis2025magnetism, joy2026disorder}). In addition to these ``old-fashioned" questions, the new setting suggests a new kind of basic question, namely: how does Berry phase physics (most specifically, a nontrivial Berry curvature) modify the nature of the WC? This question has been the subject of a number of recent theoretical works~\cite{Dong2024anomalous, soejima2024anomalous, dong2024stability, tan2024parent, zeng2024sublattice, tan2025variational, zhou2025new, soejima2025lambda, bernevig2025berry, miao2026various, zverevich2026spin}, which have proposed a variety of new or modified  WC-type phases.

The present paper approaches this question by focusing on the simplest and oldest instance of multilayer graphene: Bernal bilayer graphene (BBG), in which two monolayer graphene sheets are arranged in an untwisted AB stack. In BBG, application of a perpendicular electric field has the effect of simultaneously opening a band gap, creating a large density of states near the band edge, and distributing $\pm 2 \pi$ Berry flux over an increasingly large region of momentum space in the vicinity of the $K$ and $K'$ points in momentum space. Our goal in this paper is to consider the questions: what becomes of the WC state in BBG? Where does it exist in the parameter space of electron density, displacement field, and temperature? And how are its properties different from those of the conventional WC phase?

Our basic results can be summarized as follows. At zero displacement field, the WC state is precluded due to the vanishing band mass at low energy and the strong interband dielectric response. Instead, the WC phase exists within a regime of a large enough displacement field, low enough density, and low enough temperature. We describe, theoretically, the phase boundary of the WC state in two ways: first, by ignoring the effects of trigonal warping so that the low-energy dispersion relation is rotationally symmetric around the K and K' points; and second, by including the effects of trigonal warping, so that the low-energy band structure is split into either three or four low-energy ``pockets" or ``minivalleys." In the former case, we show that  when the displacement field is strong enough, Berry curvature drives a spontaneous orbital angular momentum transition of WC electrons, which manifests as a discontinuous jump in the magnetization as a function of the displacement field. When the effects of trigonal warping are included, however, the ground state of the WC involves a nontrivial spatial pattern of minivalley ordering. Trigonal warping also produces an unusual ``doubly-reentrant melting" of the WC phase, in which the WC phase first melts, then freezes again, then melts again with increasing density in the limit of zero temperature.

We note that this paper (which was first released on the arxiv in 2023 \cite{joy2023wignerv1}) is meant to be complimentary with another recent publication by us \cite{joy2025chiral}, which announces some of the same results included here and also examines the nontrivial spin ordering of the WC state that can result from Berry curvature. In the present paper we focus primarily on the WC phase diagram and on the orbital effects of Berry curvature, including a detailed examination of the effects of trigonal warping, which is neglected in Ref.~\cite{joy2025chiral}.

The remainder of this paper is organized as follows. In Sec.~\ref{sec:preliminaries}, we discuss the dispersion relation of BBG and present our criterion for estimating the stability of the WC phase. Section \ref{sec:MH} presents analytical derivations and numeric calculations for the case without trigonal warping and Sec.~\ref{sec:trigonal} considers the effects of trigonal warping. We close in Sec.~\ref{sec:conclusion} with a summary and a discussion of possible experimental realizations.

\section{BBG dispersion and Harmonic oscillator description of the WC state}
\label{sec:preliminaries}

Before discussing the WC state, we briefly review the electron dispersion relation in BBG and its dependence on the displacement field. BBG consists of two parallel copies of monolayer graphene, arranged so that the A-sublattice carbon atoms of one graphene sheet are vertically atop the B-sublattice atoms of the other. Therefore, the low-energy band structure can be seen as two copies of the Dirac cone that are hybridized by inter-layer tunneling (e.g., Ref.~\cite{mccann2013electronic} for a review). In the simplest description, where one accounts only for the hopping of electrons between nearest-neighboring atoms (both in-plane and vertical), the low-energy conduction and valence band energies are given by
\begin{equation}
    \varepsilon(p)=\pm\left(\frac{\gamma_{1}^{2}}{2}+v^{2}p^{2}-\gamma_{1}\left(\frac{\gamma_{1}^{2}}{4}+v^{2}p^{2}\right)^{1/2}\right)^{1/2},
\label{eq:dispersionU=0}
\end{equation}
where $\gamma_{1}$ is the interlayer tunneling amplitude and $v$ is the single-layer graphene Dirac velocity \cite{mccann2013electronic}. Equation \ref{eq:dispersionU=0} describes a set of bands that meet at a point in momentum space (the $K$ or $K'$ point) and, at low energy, disperse parabolically with the momentum $p$ relative to that point. In the remainder of this paper, except where noted explicitly, we use dimensionless units where $\hbar = v = \gamma_1 = 1$ so that all energies are in units of $\gamma_1 \approx 400$\,meV, and all densities are in units of $(\gamma_1/\hbar v)^2 \approx 4 \times 10^{13}$\,cm$^{-2}$.

\begin{figure}[htb]
\centering
\includegraphics[width=1.0 \columnwidth]{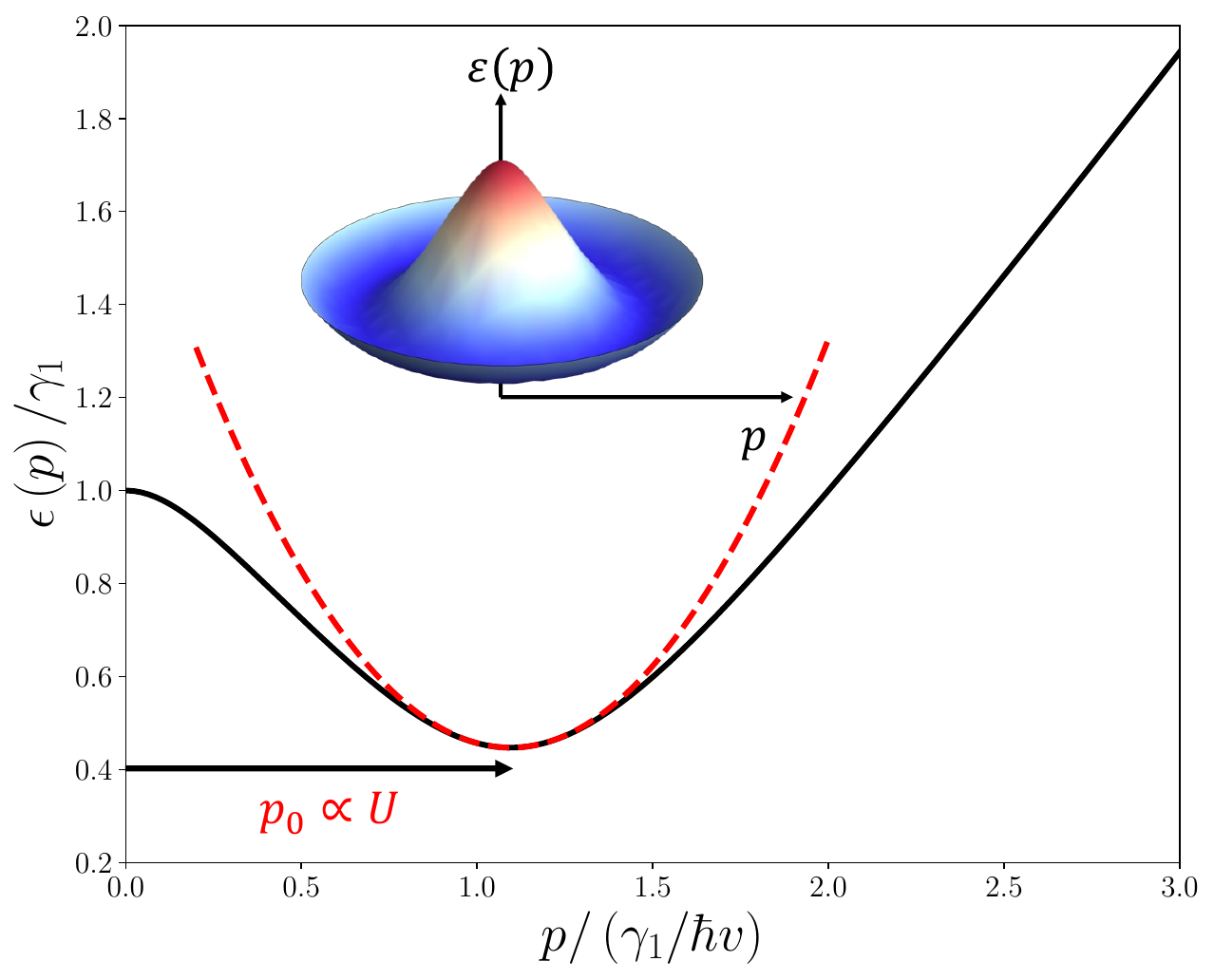}
\caption{Conduction band dispersion $\varepsilon\left(p\right)$ of BBG (Eq.~\ref{eq:disp_app}), neglecting trigonal warping, plotted for $U = 2.0$ as a function of the radial momentum $p$. The quantity $p_0$ represents the radius of the dispersion minimum. The inset shows the full "Mexican Hat" structure as a 3D plot. The red dashed line shows the approximation in Eq.~\ref{eq: MHdispersion}.}
\label{fig:DispersionCutMH}
\end{figure}

A perpendicular displacement field creates a difference $U$ in potential energy between the top and bottom layers, and the dispersion relation for the conduction band becomes:
\begin{equation}
\varepsilon\left(p\right)=\left(\frac12 + \frac{U^{2}}{4} + p^{2} - \left(\frac14 +  p^{2} \left(1 + U^2\right) \right)^{1/2}\right)^{1/2}.
\label{eq:disp_app}
\end{equation}
Equation \ref{eq:disp_app} describes a ``Mexican hat'' (MH) shape (see Fig.~\ref{fig:DispersionCutMH}), with a ring of minima located at a certain value $|\vec{p}| =p_0$. For momenta close to this ring of minima, the dispersion can be expanded as
\begin{equation}
\varepsilon\left(p\right)\simeq \frac{U}{2\sqrt{1+U^{2}}} + \frac{\left(p-p_0\right)^2}{2m},
\label{eq: MHdispersion}
\end{equation}
where 
\begin{equation}
p_{0}=\frac{U}{2}\sqrt{\frac{2+U^{2}}{1+U^{2}}}\;\;\text{and}\;\;\;
m=\frac{\left(1+U^{2}\right)^{3/2}}{2U\left(2+U^{2}\right)}.
\label{eq:p0andm}
\end{equation} 
Thus the value of $p_0$ increases approximately proportional to the interlayer potential $U$. On the other hand, the effective mass $m$ in the radial direction saturates to $1/2$ at large $U$ and diverges as $1/U$ for small $U$.

Equations \ref{eq:dispersionU=0} -- \ref{eq:p0andm} neglect the tunneling of electrons between non-neighboring atoms on opposite layers. Such skew hopping leads to ``trigonal warping'', in which the rotationally-symmetric low energy band structure described by Eq.~\ref{eq:disp_app} loses its rotational symmetry and retains only the lower, $C_3$ symmetry of the parent graphene. Specifically, the conduction band energy is given by
\begin{align}
    \begin{split}
        \varepsilon\left(p\right)&=\left[\frac{1}{2}+\frac{U^{2}}{4}+p^{2}\left(\frac{t^{2}+2}{2}\right) - \right.\\&\left.\left(2p^{3}t\cos\left(3\phi\right)+p^{2}\left(p^{2}t^{2}+U^{2}+1\right)+\frac{\left(1-p^{2}t^{2}\right)^{2}}{4}\right)^{1/2}\right]^{1/2}
    \end{split}
    \label{eq:dispersionTW}
\end{align}
where $\phi$ is the angle between the momentum vector $\vec{p}$ and the $p_x$ axis. The trigonal warping scale $t \approx 0.12$ \cite{mccann2013electronic}.

At $U = 0$, Eq.~\ref{eq:dispersionTW} implies that the parabolic band touching point is split by trigonal warping into four low energy Dirac cones: one ``central pocket'' at $p = 0$ and three ``side pockets''. As $U$ is increased from zero, all four pockets acquire an energy gap. The three side pockets move toward increasingly large $p$ and remain degenerate in energy, while the central pocket acquires both larger mass and larger energy than the side pockets. When the interlayer potential difference is higher than the energy scale associated with trigonal warping ($U > t$), the central pocket becomes a local maximum and disappears.

\begin{figure}[htb]
\centering
\includegraphics[width= 0.9 \columnwidth]{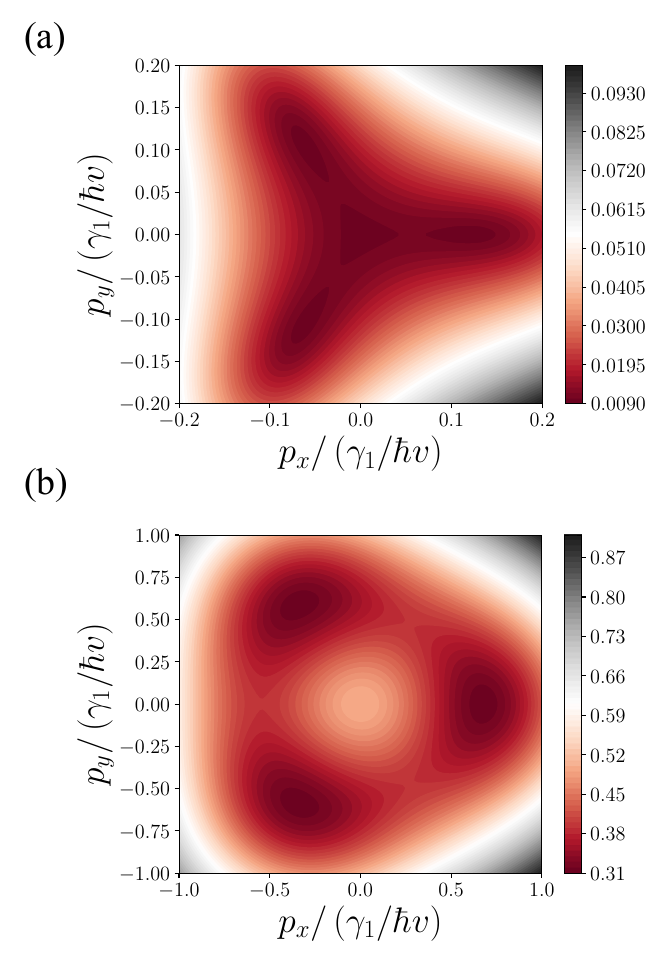}
\caption{Color plots of the conduction band dispersion $\varepsilon\left(p\right)$, including the effects of trigonal warping (see Eq.~\ref{eq:dispersionTW}), presented for (a) $U=0.02$ and (b) $U=1.0$. The energy is measured (color bar) in the units of $\gamma_1$. There is a dispersion minimum at the center, which exists only when $U<t$.  At small $U$ (specifically $U \lesssim  9t/2\sqrt{2}$), the side pockets are elongated toward the $K$ point. At large $U$ (specifically $U \gtrsim 9t/2\sqrt{2}$), the side pockets are elongated in the direction perpendicular to the $K$ point. Note the difference in energy scales at the color bar}
\label{fig:DispersionTwarp}
\end{figure}

At finite $U$, trigonal warping turns the ring of minima in the dispersion relation into three discrete minima, ``mini-valleys,'' each with the same energy. At large $U \gg t$, these mini-valleys are located at $p$ close to the value of $p_0$ given by Eq.~\ref{eq:p0andm} and at $\phi = 0, \pm 2\pi/3$. These different limits are illustrated in Fig.~\ref{fig:DispersionTwarp}.

\begin{figure}[htb]
\centering
\includegraphics[width=1.0 \columnwidth]{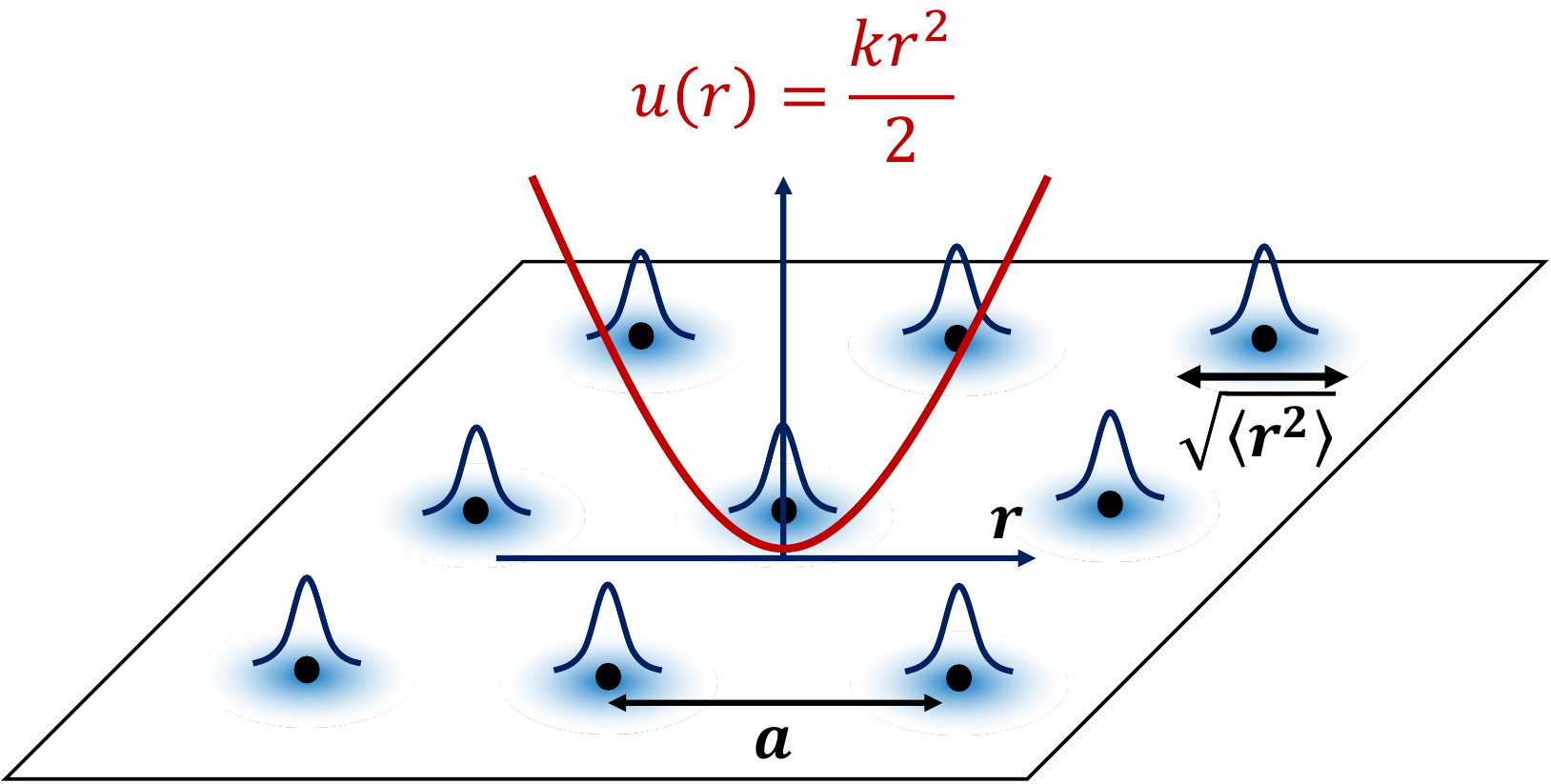}\\
\caption{Schematic representation of the WC lattice, showing the lattice constant $a$ of the Wigner lattice and the electron wave packet width $\sqrt{\langle r^2 \rangle}$. The red curve denotes the harmonic confinement experienced by the corresponding electron, and the blue-shaded regions represent the HO ground state wavepackets.}
\label{fig:WC_Lattice}
\end{figure}

We now review the general criterion for the stability of the WC phase. The conventional semiclassical model for the WC in two dimensions describes individual electrons as being localized to points in a triangular lattice that minimizes the classical electrostatic energy (see Fig.~\ref{fig:WC_Lattice}). In this arrangement, each electron resides in a local minimum of the electrostatic potential created by all other electrons. Thus, deep in the WC phase, the primary contribution to the energy per electron is equal to the classical electrostatic energy. To understand the lowest-order quantum correction, consider that each electron can be described as a harmonic oscillator (HO) residing in a locally parabolic potential whose strength is determined by the Coulomb interaction \cite{mahan1990many, Flambaum1999Spin}. One can therefore estimate the lowest-order quantum correction to the energy per electron as the ground state energy of the two-dimensional HO. In the conventional WC state, the HO description correctly gives the lowest-order quantum correction to the energy with an accuracy better than 10$\%$ \cite{mahan1990many}.

The HO picture also provides a straightforward way of estimating the critical density associated with the quantum melting of the WC. For a conventional particle with mass $m$ in a confining potential $u(r) = k r^2/2$, where $r$ is the distance from the potential minimum, the HO ground state wavefunction has a characteristic frequency $\omega = \sqrt{k/m}$ and root-mean-squared radius $\sqrt{\langle r^2 \rangle} = \sqrt{\hbar/m \omega}$. Melting is associated with the ratio $\eta = \sqrt{\langle r^2 \rangle}/a$, where $a = (\sqrt{3} n / 2)^{-1/2}$ is the lattice constant of the Wigner lattice, becoming larger than a critical value $\eta_c$ (the Lindemann criterion). Empirically, $\eta_c$ is universally in the range $0.20 - 0.25$ \cite{babadi2013universal, AstrakharchikQuantum2007}. While making a precise determination of the critical density associated with WC melting is, in general, very difficult, here we use the Lindemann criterion (with $\eta_c = 0.23$) as a simple proxy to estimate the critical density $n_c$ associated with the melting of the WC state \cite{Joy2022Wigner}:
\be 
\sqrt{\langle \hat{\vec{r}}^2 \rangle } = \eta_c \left( \frac{2}{\sqrt{3} n_c} \right)^{1/2} .
\label{eq:Lindemann}
\ee 
In this way, our discussion of the WC state is reduced to solving a single-particle problem: that of a single-particle harmonic oscillator in a confining potential whose strength is determined by electron-electron interactions. Below we provide more discussion of the limitations of the HO picture for determining the ground state of the WC. We note here only that using Eq.~(\ref{eq:Lindemann}) to predict the location of the conventional WC-FL transition gives a critical $r_s \approx 34$, as compared to the value $r_s \approx 30 - 40$ predicted by quantum Monte Carlo calculations \cite{drummond2009phase, azadi2024quantum} and $r_s \approx 1-4$ predicted by Hartree-Fock \cite{trail2003unrestricted, bernu2011hartree}. 

Let us briefly recall the structure of energy levels of the conventional 2D HO. The eigenstate of the 2D HO can be labeled by the principle and azimuthal quantum numbers $(n, \ell)$, and the eigenenergies are $E_n = (n+1)\hbar \omega$, with $n$ a non-negative integer. The $n^\text{th}$ energy level is  $(n+1)$-fold degenerate (ignoring spin degeneracy). For odd $n$, the allowed values of $\ell$ are odd, $\ell \in (-n, -n+2, ...,n)$, whereas for even $n$, only even values of $\ell$ are allowed, $\ell \in (-n, -n+2,...,0,...,n)$. It should be noted that the $(n, \ell) = (0,0)$ state remains the ground state even when an external magnetic field is applied \cite{faddeev2004va}. A large magnetic field can bring states with finite $\ell$ close to the ground state (the energy difference between $\ell\neq0$ and $\ell=0$ is given by $\Delta E_{\ell}\simeq |\ell| \hbar\omega^{2}/2\omega_{c}$, where $\omega_{c}=eB/m$ is the cyclotron frequency), but the ground state always has $\ell = 0$.

When considering electrons with an arbitrary dispersion relation $\e(p)$ in a parabolic confining potential, it is simplest to write the Hamiltonian in momentum space:
\begin{equation}
H = \e(p) + \frac12 k \hat{\vec{r}}^2,
\label{eq:H_eff_1}
\end{equation}
where $\hat{\vec{r}}$ is the position operator. For a band that has nonzero Berry curvature, one can arrive at the effective Hamiltonian by writing the position operator $\hat{\vec{r}}$ in momentum space and then projecting the resulting Hamiltonian to the band of interest. The resulting Hamiltonian (see Appendix \ref{sec:HO_BC} for details) is given by 
\begin{equation}
H = \varepsilon\left(\vec{p}\right)+\frac{k}{2}\left(\dot{\iota}\vec{\nabla}_{\vec{p}} + \vec{A}\left(\vec{p}\right)\right)^{2},
\label{eq:H_eff_2}
\end{equation}
where $\vec{A}\left(\vec{p}\right)$ is the Berry connection of the band of interest \cite{Price2014Quantum, Berceanu2016Momentum, Karplus1954Hall, Lapa2019Semiclassical}. Comparing Eq.~\ref{eq:H_eff_2} to the usual HO Hamiltonian written in position space, one can say that the dispersion relation $\e(p)$ acts like a scalar potential in momentum space, while the Berry connection $\vec{A}(\vec{p})$ acts like a magnetic vector potential.

In the following sections, we compute the low-energy eigenstates of Eq.~\ref{eq:H_eff_2} for a given strength $k$ of the confining potential. The value of $k$ is determined by the interactions between electrons in the Wigner lattice and is, therefore, a function of the electron density, defined as follows. If one takes the origin $\vec{r} = 0$ to be a site of the Wigner lattice, then the potential $u(\vec{r})$ experienced by the electron near the origin is $u(\vec{r}) = \sum_{i\neq0}V\left(\left|\vec{r} - \vec{R}_{i}\right|\right)$, where $i$ indexes the sites of the Wigner lattice, $V(r)$ is the Coulomb interaction, $\vec{R}_i$ denotes a lattice vector of the Wigner lattice, and $\vec{r}$ is the position of the electron near the origin. Here we have assumed that $|\vec{r}|$ is much smaller than the Wigner lattice constant so that other electrons in the Wigner lattice can be treated as having a fixed position. The value of $k$ can then be found by Taylor expanding $u(\vec{r})$ to the second order in $r$, which gives \cite{Joy2022Wigner}
\begin{equation}
k = \frac{1}{2}\sum_{i\neq0}\left[V''\left(\left|\vec{R}_{i}\right|\right)+\frac{V'\left(\left|\vec{R}_{i}\right|\right)}{\left|\vec{R}_{i}\right|}\right].
\label{eq: ho}
\end{equation}
In general, $V(r)$ should be taken as the screened Coulomb interaction, which we discuss in Sec.~\ref{sec:dielectric}. 

From our calculation of the ground state of Eq.~\ref{eq:H_eff_2}, we can estimate whether the WC phase is stable at a particular value of the electron density and displacement field, and we can also assess whether the ground state of the WC has finite or zero orbital angular momentum. 

\section{Wigner crystal phase with a Mexican hat dispersion}
\label{sec:MH}

In this section, we consider the properties of the WC state under the assumption that the trigonal warping term in the dispersion relation can be neglected so that the dispersion relation is rotationally symmetric around each valley. We show in Sec.~\ref{sec:trigonal} that including the effects of trigonal warping leaves some of the features of the phase diagram intact but also changes the nature of the WC state at large $U$. The case of no trigonal warping offers the advantage that most of its properties can be solved analytically, and it provides a clear theoretical example of a WC state with spontaneous orbital polarization.

We mention that previous authors have considered the fate of the WC state when the dispersion relation has a MH shape, both in the context of electron systems with spin-orbit coupling \cite{berg_electronic_2012, Silvestrov2014Wigner} and in the context of BBG \cite{Silvestrov2017Wigner}. These papers argue that the ground state of the WC at low density involves a spontaneous breaking of the Wigner lattice symmetry, with electrons having spatially elongated wavefunctions in real space that are condensed along one region of the minimal ring of the dispersion relation in momentum space. Our analysis here, based on the HO description, suggests instead that at low-density electron wavefunctions remain essentially rotationally symmetric. We discuss this difference in more detail in Sec.~\ref{sec:symmetrybreaking}, including the limitations of the HO description. However, as we demonstrate below, incorporating the influence of Berry curvature and trigonal warping leads to qualitative changes to the properties of the WC phase that facilitate symmetry breaking similar to the kind suggested in Refs.~\cite{berg_electronic_2012, Silvestrov2014Wigner, Silvestrov2017Wigner, calvera2022pseudo}.

\subsection{Without Berry curvature}

In the absence of Berry curvature, the Schrödinger equation (SE) in momentum space for the Harmonic oscillator potential is given by
\begin{align}
    \begin{split}
        \left[\varepsilon\left(p\right)-\frac{k}{2}\left(\frac{\partial}{\partial p}\left(p\frac{\partial}{\partial p}\right)+\frac{1}{p^{2}}\frac{\partial^{2}}{\partial\phi^{2}}\right)\right]\psi\left(\vec{p}\right)=E\psi\left(\vec{p}\right).
    \end{split}
    \label{eq:SE}
\end{align}
Due to the rotational symmetry of the problem, the eigenstates of the Hamiltonian can be written using the separation of variables as $\psi\left(\vec{p}\right)=f\left(p\right) \exp(\dot{\iota}\ell\phi)$, where $\ell$ is the angular momentum quantum number (throughout this paper we use $\phi$ to denote the azimuthal angle in the momentum space, while $\theta$ denotes the azimuthal angle in real space). We can further convert the above SE into an effective one-dimensional SE by defining ${f(p) \equiv g(p)/\sqrt{p}}$ and introducing a dimensionless momentum ${u \equiv p/\sigma}$, ${u_0 \equiv p_0/\sigma}$, with $\sigma = \sqrt{ m\omega}$ being the characteristic momentum of a HO. (As above, $\omega = \sqrt{k/m}$ is the characteristic HO frequency.) The resulting SE becomes
\begin{equation}
\left[-\frac{d^{2}}{du^{2}}+\frac{1}{u^{2}}\left(-\frac{1}{4}+\ell^{2}\right)+\left(u-u_{0}\right)^{2}\right] g(u) =\epsilon g(u),
\label{eq: 1DSE}
\end{equation}
where we have defined $\epsilon\equiv E\big/(\omega/2)$, and we have used Eq.~\ref{eq:disp_app} to approximate the dispersion of $\varepsilon(p)$. 

As the harmonic confinement becomes asymptotically weak (i.e., at very low electron density), the value of the constant $u_0$ becomes large, and the eigenstates $g(u)$ are concentrated around $u = u_0$. Expanding the SE in terms of $x \equiv (u - u_0) \ll u_0$ gives
\begin{equation}
\left(-\frac{d^{2}}{dx^{2}} + x^{2} \right) g(x) = 
\left(\epsilon-\frac{1}{u_{0}^{2}}\left(-\frac{1}{4}+\ell^{2}\right)\right)g(x),
\end{equation}
which is precisely the usual SE for a 1D HO. (This mapping between a 2D bound state for an electron with a Mexican Hat dispersion and an equivalent 1D problem has been pointed out by previous authors, particularly in the context of a hydrogen-like bound state \cite{Chaplik2006bound, Skinner2014Bound, Skinner2019properties}.)
We can now read off the low-energy spectrum as
\begin{equation}
    E_{\ell}\simeq\frac{\omega}{2}+\frac{\omega\ell^{2}}{2u_{0}^{2}}-\frac{\omega}{8u_{0}^{2}}.
    \label{eq:spectrum_wo}
\end{equation}
Appendix \ref{sec: MHHO} gives more details about this derivation.

Notice that Eq.~\ref{eq:spectrum_wo} is qualitatively different from the energy spectrum of the conventional two-dimensional HO. Most notably, when the principal quantum number $n=0$, there remains a large number of nearly degenerate angular momentum states $\ell$ with energy level spacing that is proportional to $1/u_0^{2}$. 

\begin{figure}[htb]
\centering
\includegraphics[width=1.0 \columnwidth]{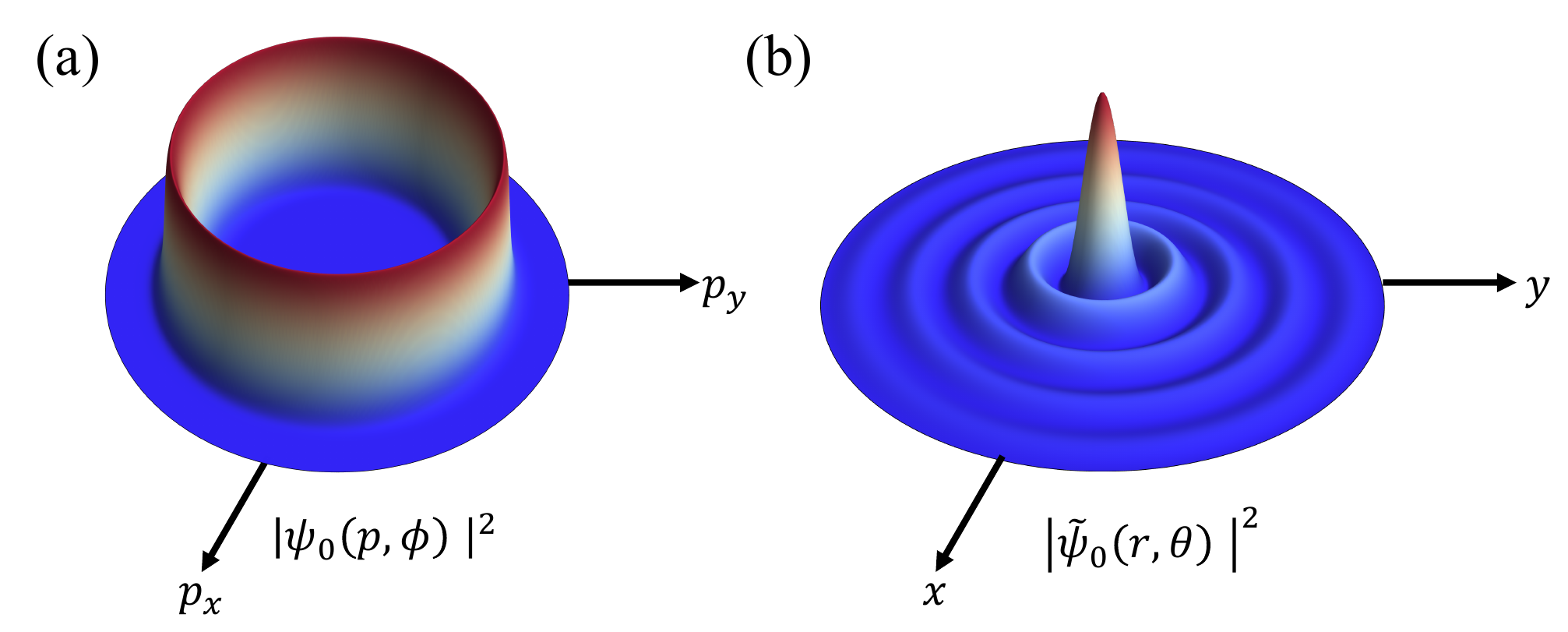}\\
\caption{The probability density of the ground state wavefunction is plotted in (a) the momentum space (Eq.~\ref{eq:ms_wf}) and (b) the real space (Eq.~\ref{eq:rs_wf}).}
\label{fig: wf}
\end{figure}

The normalized wavefunctions take the following form in momentum space:
\begin{equation}
\psi_{\ell}\left(p,\phi\right)=\left(\frac{2\sqrt{\pi}}{\sigma}\right)^{1/2}\frac{\exp\left[-\frac{\left(p-p_{0}\right)^{2}}{2\sigma^{2}}\right]}{\sqrt{p}}\:\exp\left[\dot{\iota}\ell\phi\right].
\label{eq:ms_wf}
\end{equation}
In real space, these eigenstates are given by 
\begin{equation}
    \tilde{\psi}_{\ell}\left(r,\theta\right)=\left(\frac{p_{0}\sigma}{\sqrt{\pi}}\right)^{1/2}J_{\ell}\left(p_{0}r\right)\exp\left[-\frac{\sigma^{2}r^{2}}{2}\right]\exp\left[\dot{\iota}\ell\theta\right],
\label{eq:rs_wf}
\end{equation}
where $J_\ell$ is the $\ell^{th}$ order Bessel function of the first kind. The probability distribution corresponding to the ground state wavefunction in momentum space and real space are plotted in Fig. \ref{fig: wf}. Note that the wavefunction exhibits fast oscillations in real space with wavelength $\sim 1/p_0$ and many nodes of density even in the ground state. This violation of the usual rule that ``ground state wavefunctions don't have nodes'' arises from the unique ring of minima in the dispersion relation.

The width of the wavefunction in real space can be calculated to be
\begin{equation}
\left\langle \hat{r}^{2}\right\rangle_{\ell} \simeq\frac{1}{\sigma ^2}\left(\frac{1}{2}+\frac{\ell^{2}}{u_{0}^{2}}-\frac{1}{4u_{0}^{2}}\right).
\label{eq: nclargeU}
\end{equation}
We can use this result, together with the Lindemann criterion (Eq.~\ref{eq:Lindemann}) to estimate the critical density associated with the WC phase. In the limit of large $U \gg 1$, where the mass approaches $m \simeq 1/2$ (see Eq.~\ref{eq:p0andm}) and dielectric screening is unimportant (see Sec.~\ref{sec:dielectric} below), this calculation yields 
\be 
n_c \approx 2.1\times10^{-4}.
\ee 
(We will show in the following section that including the effects of trigonal warping leads to a lower numerical estimate of $n_c$.)

\subsection{Possibility of spontaneous rotational symmetry breaking of the electron wavefunction}
\label{sec:symmetrybreaking}

So far, we have discussed the WC state using the HO description. Under this description, the lowest energy state of the electron is the one derived in the previous subsection, which has rotational symmetry. Other proposals for the WC state with a MH dispersion have suggested that the electron wavefunction at each site of the Wigner lattice condenses along one point of the ring of minima in the dispersion so that, effectively, the electron has a very heavy mass in one direction and a light mass in the perpendicular direction, leading to a highly anisotropic wavefunction \cite{berg_electronic_2012, Silvestrov2014Wigner, Silvestrov2017Wigner}. Within the HO description, the radially symmetric electron wavefunction that we have derived has a lower energy than these spatially elongated wavefunctions by an amount $\sim \omega (\sigma/p_0)^2$. 

But it is important to note that the HO description alone cannot accurately distinguish between these candidate ground states for the WC because it fails to capture the effects of correlations in the quantum fluctuations between different electrons in the Wigner lattice. To correctly calculate the lowest-order quantum correction to the WC energy, one should, in general, calculate the dispersion relation $\omega(\vec{q})$ of WC phonon modes; the ground state has one quantum of energy $\omega(\vec{q})$ for each possible phonon mode $\vec{q}$ \cite{Bonsall1977Some}. The HO description provides an overestimate of the WC energy because it does not account for these correlated quantum fluctuations. Alternatively, one can say that the HO energy we derive represents the highest frequency in the WC phonon dispersion -- it is the ``Einstein phonon'' limit. In the conventional WC, the HO description is fairly accurate: it gives an estimate for the numerical value of the quantum correction to the WC energy at low density that is only $\approx 9\%$ larger than the value given by the phonon calculation \cite{mahan1990many}.\footnote{It is worth noting that the HO description gives an energy that is equivalent to calculating the Hartree energy of a trial wavefunction that consists of Gaussian wave packets at every site of the Wigner lattice \cite{skinner2013effect, skinner2016interlayer, Maki1983Static, Chitra2005Zero}.}

Unfortunately, in our case, the phonon calculation is difficult to carry out due to the non-monotonic dependence of the electron kinetic energy on momentum, and it cannot be reproduced by a simple extension of the canonical procedure \cite{Bonsall1977Some}. We are therefore unable to estimate in a precise way the quantum correction to the WC energy, which would adjudicate between different candidate ground states. However, we speculate that for the MH dispersion the rotationally symmetric state we are describing is lower in energy than the symmetry-broken states described by Ref.~\cite{berg_electronic_2012, Silvestrov2014Wigner, Silvestrov2017Wigner}, since in the former case the maximum frequency in the WC phonon dispersion is lower, and the electron wavefunctions are more compact in real space, leading to reduced interaction in general.

However, for real BBG, this discussion is largely irrelevant because the presence of trigonal warping breaks the rotational symmetry of the dispersion and leads to anisotropic mini-valleys that naturally yield spatially elongated wavefunctions. So, in fact, the WC state in real BBG at low density does consist of a pattern of spatially elongated wave packets with nontrivial mini-valley ordering on the WC lattice \cite{calvera2022pseudo}, as we discuss in the following section.

\subsection{With Berry curvature}
\label{sec:MHBC}

We now return to the HO description and consider the effects of Berry curvature on the WC state by including the Berry curvature in the effective single-particle  Schrödinger equation. It should be noted that in BBG, the Berry curvature takes opposite signs in the $K$ and $K'$ valleys. Here, for simplicity, we focus primarily on the $K$ valley, where the Berry curvature is positive. The details of the derivation from this subsection can be found in Appendix \ref{sec:MHHOBC}.

One can include the effects of Berry curvature by noticing that the Berry connection enters the effective single-electron Hamiltonian (Eq.~\ref{eq:H_eff_2}) via a minimal substitution to the position operator written in momentum space. The Berry connection  $\vec{A}$ is particularly straightforward to write in the Coulomb gauge when the Berry curvature $\Omega\left(p\right)$ is radially symmetric. In this case
\begin{equation}
\vec{A}=\frac{\tilde{\Phi}\left(p\right)}{p}\hat{\phi},
\end{equation}
where $\tilde{\Phi}$ is the fraction of the total $2 \pi$ Berry flux through a disk in momentum space of radius $p$:
\begin{equation}
\tilde{\Phi}\left(p\right)=\int_{0}^{p}dp'\,p'\,\Omega\left(p'\right).
\end{equation}
The minimal substitution process of adding the Berry connection amounts to modifying the azimuthal part of the gradient operator in the following way.
\begin{equation}
\frac{\dot{\iota}}{p}\frac{\partial}{\partial\phi_{p}}\longrightarrow\frac{\dot{\iota}}{p}\frac{\partial}{\partial\phi_{p}}+\frac{\tilde{\Phi}\left(p\right)}{p}.
\end{equation}
The effective one-dimensional SE in this situation can be shown to be
\begin{equation}
\left[-\frac{d^{2}}{du^{2}}+\frac{1}{u^{2}}\left(-\frac{1}{4}+\left(-\ell+\tilde{\Phi}\left(u\sigma\right)\right)^{2}\right)+\left(u-u_{0}\right)^{2}\right]g(u) = \epsilon g(u).
\label{eq:1DSE_with}
\end{equation}
Taylor expanding this equation for small $x = u - u_0$, as in the previous subsection, gives
\begin{equation}
-\frac{d^{2}g}{dx^{2}} + x^{2}g(x) = \left(\epsilon-\frac{1}{u_{0}^{2}}\left(-\frac{1}{4}+\left(\ell-\tilde{\Phi}\left(p_{0}\right)\right)^{2}\right)\right) g(x) ,
\end{equation}
so that the eigenenergies are given by
\begin{equation}
E_\ell \simeq \frac{\omega}{2}+\frac{\left(\ell-\tilde{\Phi}\left(p_{0}\right)\right)^{2}\sigma^2\omega}{p_{0}^{2}}-\frac{\sigma^2\omega}{4p_{0}^{2}}.
\label{eq:spectrum_wi}
\end{equation}
(Equation \ref{eq:spectrum_wi} was previously derived in Ref.~\cite{Price2015Artificial} in the context of a spin-1/2 particle with Rashba spin-orbit coupling in harmonic confinement.)

Equation \ref{eq:spectrum_wi} implies that the Berry flux enclosed by the wavefunction in momentum space plays a crucial role in determining the electron's energy spectrum.\footnote{Notice that the standard semiclassical treatment, in which one treats the wavefunction as a compact wave packet localized around a specific point $\vec{p}$ in momentum space \cite{Xiao2010Berry}, fails to capture the effect we are describing here. In this standard treatment, the Berry curvature $\Omega$ enters only through the value $\Omega(\vec{p})$, whereas here, due to the ring of minima in the band edge, the electron state crucially depends on the \emph{flux} of Berry curvature through the wavefunction.}
In particular, if $\tilde{\Phi}(p_0) > 1/2$, the ground state has $\ell = 1$ rather than $\ell = 0$. Thus, a transition from zero angular momentum to a finite angular momentum of electrons in the WC can be induced by increasing the displacement field $U$, which increases $p_0$ and thereby encloses more flux within the wavefunction. (In the $K'$ valley, the transition is from $\ell = 0$ to $\ell = -1$.)

The critical field associated with this transition to finite angular momentum can be found by the condition: 
\begin{equation}
\int_{0}^{p_{0}\left(U_{c}\right)}dp' p' \Omega\left(p'\right) =\frac{1}{2}.
\end{equation}
(In the hypothetical case where the Berry curvature is constant as a function of $p$, this condition simplifies to $\Omega\left(p_0\right) p_0^2 =1$.) Evaluating this condition numerically gives (again, ignoring the effects of trigonal warping), 
\be 
U_c \approx 1.07.
\label{eq: Uc}
\ee 
\subsection{Magnetization}
\label{sec:magnetization}

A sudden change in angular momentum leads to observable effects in the magnetization of the WC state. The magnetization operator can be expressed as
\begin{equation}
\hat{M}_z=\frac{e}{2}\left(\vec{v}_{\vec{p}}\times\vec{r}\right),
\end{equation}
where $\vec{v}_{\vec{p}}=\vec{\nabla}_{\vec{p}}\varepsilon(\vec{p})$ is the velocity operator. The expectation value of the magnetization can be written as
\begin{equation}
\left\langle \hat{M}_z\right\rangle =\frac{e}{2}\int \frac{d^2\vec{p}}{\left(2\pi\right)^2}\,v_{p}\left(-\ell+\tilde{\Phi}\left(p\right)\right)\left|f\left(p\right)\right|^{2}.
\label{eq:mag_full}
\end{equation}
The magnetization has two contributions, one from the angular momentum of the wavefunction and the other from the underlying Berry curvature. Tuning the value of $U$ across the $\ell=0$ to $\ell=1$ transition results in a jump in the magnetization. In the limit of $u_0\gg1$, we can evaluate $\langle \hat{M}_{z}\rangle$ analytically as
\begin{equation}
\left\langle \hat{M}_z\right\rangle \simeq\frac{e}{2m}\left(\frac{\sigma^{2}}{2p_{0}^{2}}\right)\left[\ell-\tilde{\Phi}\left(p_{0}\right)+\Omega\left(p_{0}\right)p_{0}^{2}\right].
\label{eq:mag}
\end{equation}
 
Note that the magnetization in this system, for a given angular momentum, is weaker than the usual Bohr magneton $e \hbar / 2m$ by a factor $\sigma^2/2p_0^2 \ll 1$.  One can think that this large suppression of the magnetization arises because the group velocity is very small for momenta $p$ close to $p_0$ due to the flattened MH dispersion, so the electron current is weak for a given angular momentum. Note also that the sign of the magnetization is opposite to the case of the usual parabolic dispersion, which implies that the sign of $\left\langle \hat{M}_z\right\rangle$ for fixed $\ell$ is inverted as $p_0$ is increased from zero. This inversion of the sign of the magnetization is discussed further in Appendix \ref{sec: OrbMag}.

Equation \ref{eq:mag} implies that at the critical field $U_c$, which marks the transition between the $\ell=0$ and $\ell=1$ phases, the magnetization has a jump of magnitude $(e/2m)\left(\sigma^{2}/2p_{0}^{2}\right)$.

\subsection{Estimate of exchange integral: ordering temperature}
\label{sec: exchange_order}

Each electron in the WC phase has an Ising-like degree of freedom associated with the valley ($K$ or $K'$). Since the sign of the Berry curvature is opposite for the two valleys, ordering to a specific valley is concomitant with an orbital ferromagnetic transition in the $|\ell| = 1$ phase (ordering to $\ell = 1$ in one valley and $\ell = -1$ in the opposite valley). At sufficiently low temperatures, valley ordering is driven by the exchange interaction between neighboring electrons. One can arrive at a naive estimate for the temperature scale associated with valley ordering by calculating the usual exchange integral \cite{landau2013quantum} 
\begin{align}
    \begin{split}
        J_{ab}&=\int d\vec{r_{1}}\int d\vec{r_{2}}\;\left[\psi^{*}\left(\vec{r}_{1}\right)\psi^{*}\left(\vec{r}_{2}-\vec{R}\right)\right. \times \\&\left.V\left(\vec{r}_{1}-\vec{r}_{2}\right)\psi\left(\vec{r}_{2}\right)\psi\left(\vec{r}_{1}-\vec{R}\right)\right],
    \end{split}
\end{align}
where $J_{ab}$ is the exchange energy between two electrons (nominally described by single-electron wavefunctions $\psi_{\ell}\left(r\right)$): one centered at the origin and the other at a neighboring site $\vec{R}$ of the Wigner lattice. We emphasize that $J_{ab}$ is only a lower bound for the actual exchange interaction since the presence of higher-order ring exchange processes generally leads to an exponential enhancement of the exchange in the WC and favors ferromagnetic ordering \cite{Roger1984Multiple, chakravarty1999wigner, Katano2000Multiple, Kim2022interstitial}. Evaluating $J_{ab}$ using the real space wavefunction in Eq.~\ref{eq:rs_wf} (see Appendix \ref{sec: exchange} for the derivation), one arrives at
\begin{equation}
J_{ab}\approx\sqrt{\frac{8}{\pi}}\left(\frac{1}{R\sigma}\right)^{2}\left(\frac{e^{2}\sigma}{\epsilon_{r}}\right)\left(\cos\left(p_{0}R-\ell\pi-\frac{\pi}{2}\right)\right)^{2}\exp\left[-\frac{\sigma^{2}R^{2}}{2}\right].
\label{eq: Jabcos}
\end{equation}
Using this formula at the critical density (presented below) associated with melting of the WC in the $|\ell| = 1$ phase gives a value of $J_{ab}$ that is of the order $\approx 10$\,mK. (Which, as mentioned above, should be considered a lower bound for the ordering temperature.) The ordering temperature decreases exponentially as the density is reduced. Note also that  Eq.~\ref{eq: Jabcos} implies an exchange temperature that oscillates as a function of the density $n \sim 1/R^2$.

\subsection{Comment on the dielectric function}
\label{sec:dielectric}

When the interlayer potential $U$ is small, the gap between conduction and valence bands is also small, so the electron-electron interaction is significantly modified by screening by virtual interband excitations. This effect is captured by the static polarization function $\Pi(q)$, such that the Fourier-transformed Coulomb interaction is \cite{mahan1990many}
\be 
V(q) = \frac{V_0(q)}{1 - \Pi(q) V_0(q)},
\ee 
where $V_0(q) = 2 \pi e^2/(\epsilon_r q)$ is the unscreened Coulomb interaction. For gapped BBG, $\Pi(q)$ has the following asymptotic behaviors \cite{Silvestrov2017Wigner}: 

\begin{equation}
\Pi\left(q\right)=\left\{ \begin{array}{cc}
-q/4, & q\gg1\\
-\ln(4)/\pi, & 1\gg q\gg\sqrt{U}\\
-4q^{2}/3\pi U, & q \ll \sqrt{U}
\end{array}\right. .
\label{eq: Pi}
\end{equation}
As we show below, the WC state melts at densities $n \ll U$, so that the relevant momentum scale $q \sim n^{1/2}$ corresponds to the final regime in Eq.~\ref{eq: Pi}. Thus, we can use 
the screened interaction:
\begin{equation}
V\left(q\right)=\frac{2\pi e^{2}}{\epsilon_{r}(q+ q^{2}/q_{0} )},
\end{equation}
where $q_{0} = 3U/(8e^2)$. For the numerical estimates below, we have used $\epsilon_{r}\approx5$, corresponding to a hexagonal boron nitride substrate.

Notice that at sufficiently large $q$ (low electron density), the Coulomb potential $V(q) \simeq 3U/(8\epsilon_r q^2)$ corresponds to a logarithmic dependence of $V(r)$ on the spatial distance $r$. This modified Coulomb interaction significantly modifies the critical density associated with WC melting \cite{Joy2022Wigner} whenever $n \gg U^2$. In the limit $U \ll 1$ and $U \ll q \ll \sqrt{U}$, the relevant electron dispersion is quartic, given by $\varepsilon(p)\approx U/2 + p^4/U$ \cite{Silvestrov2017Wigner}. Using the screened interaction for $V(q)$, and calculating the confining potential strength via Eq.~\ref{eq: ho}, one can show that $k=3\pi U n /8$ \cite{Joy2022Wigner}. The Lindemann criterion for melting then gives a critical density
\begin{equation}
n_c = C_2 \times U,
\label{eq: ncSmallU}
\end{equation}
where $C_2\approx 173$. Thus, unlike in the conventional WC problem, in BBG, the critical density vanishes in the limit $U \rightarrow 0$. That is, interlayer dielectric response precludes the formation of a WC state unless a displacement field is applied that provides an energy gap for virtual electron-hole pairs.
\subsection{Numerical results}
\label{sec:MHnumerics}

We can map out the full phase diagram for the WC phase by implementing a numerical solution of the effective one-dimensional radial SE (see Eq.~\ref{eq:1DSE_with}) for a given density $n$ and interlayer potential $U$. The stability of the WC phase is estimated by numerically calculating $\langle r^2 \rangle$ for the ground state wavefunction and using the Lindemann criterion (Eq.~\ref{eq:Lindemann}). Whether the ground state has $\ell = 0$ or $\ell = 1$ is estimated by comparing the energies of the two solutions. We use the Numerov algorithm \cite{hamming2012numerical} to solve the SE numerically.

\begin{figure}[htb]
\centering
\includegraphics[width=1.0 \columnwidth]{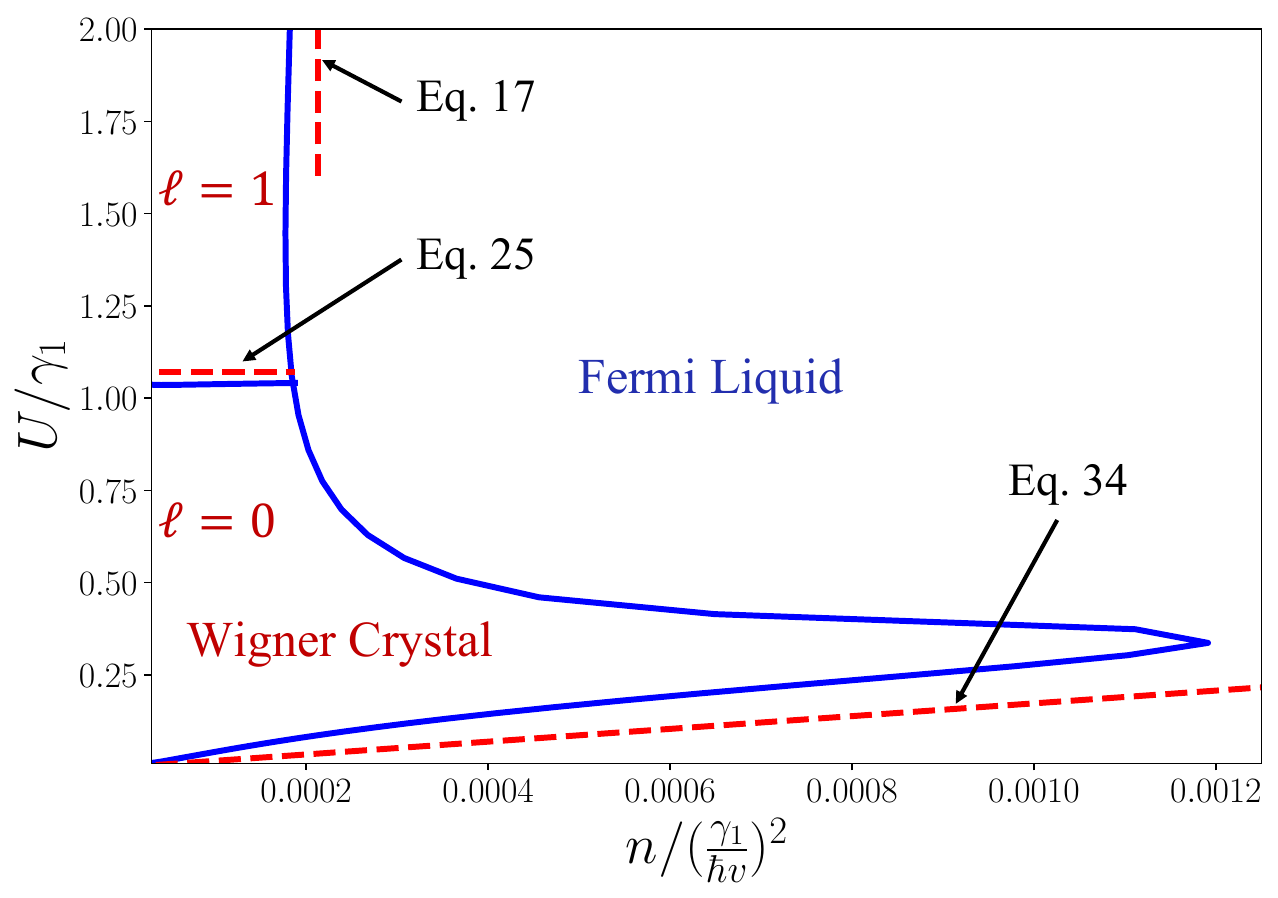}
\caption{The phase diagram of the low-density electron gas in BBG in the space of electron density $(n)$ and displacement field $(U)$. The blue lines represent the phase boundary between Wigner Crystal (WC) and Fermi Liquid (FL), with the critical value of $U_c$ associated with the orbital polarization calculated numerically. The dashed red lines correspond to the analytically calculated melting densities $n_c$ in both the small and large $U$ limits (see Eqs.~\ref{eq: nclargeU} and \ref{eq: ncSmallU}), as well as the critical $U_c$ (see Eq.~\ref{eq: Uc}), which distinguishes between WC states with angular momentum $\ell=0$ and $\ell=1$.} 
\label{fig: PD}
\end{figure}

The numerically calculated phase diagram is presented in Fig.~\ref{fig: PD}. The red dashed lines indicate the critical values $n_c$ and $U_c$ associated with melting and orbital polarization of the WC state that were derived analytically in the previous subsections. The extension of the WC state toward large $n$ at small $U$ is associated with the effective mass at the bottom of the band becoming very heavy as $U$ is reduced (Eq.~\ref{eq:p0andm}). The disappearance of the WC state as $U \rightarrow 0$ arises due to dielectric screening, which truncates the long-ranged part of the Coulomb interaction when the band gap vanishes.

\begin{figure}[htb]
\centering
\includegraphics[width=1.0 \columnwidth]{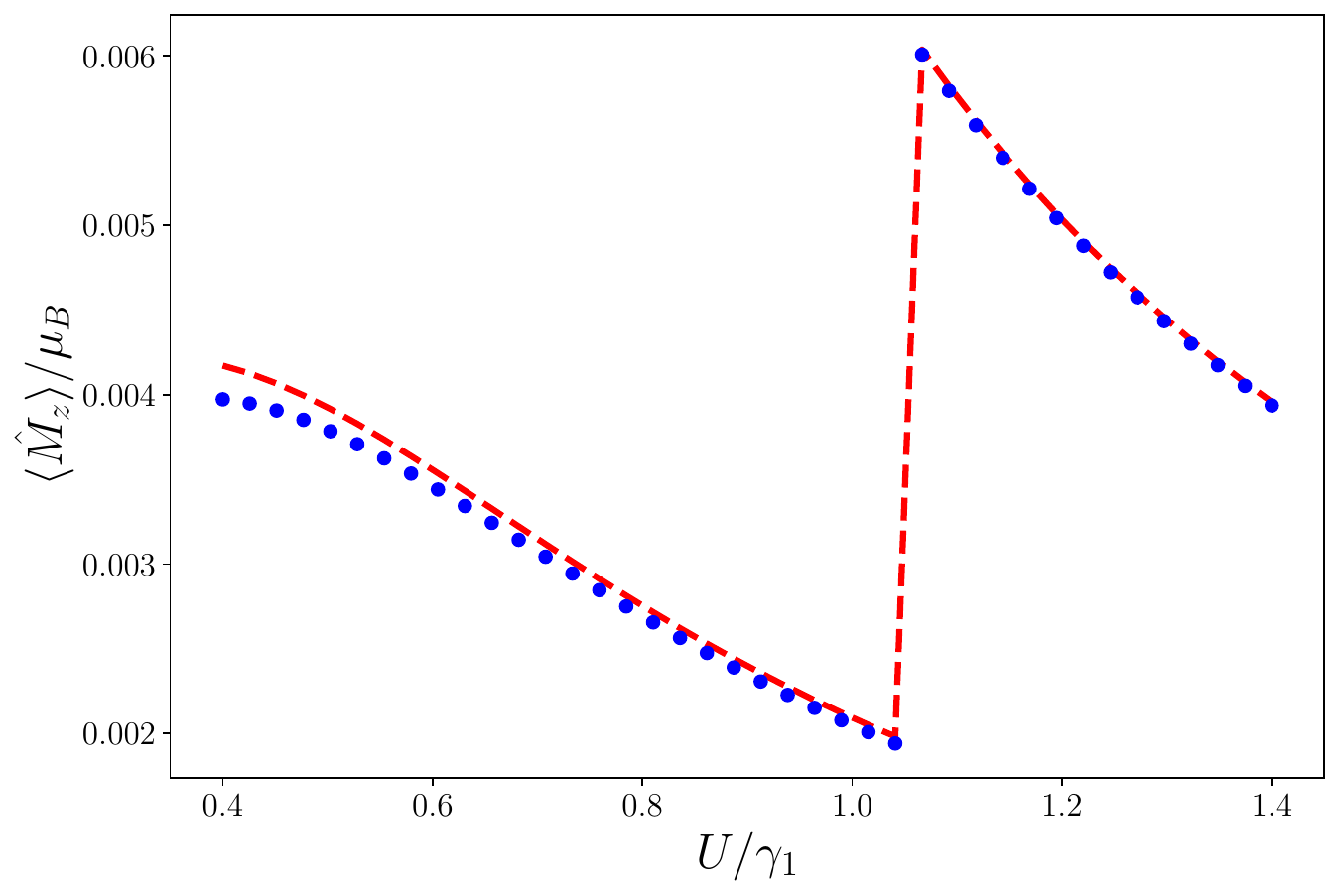}
\caption{Magnetization per electron plotted as a function of displacement field $U$ at $n=2\times10^{-4}$. The blue dots represent the numerical results, and the red dashed line corresponds to the analytical result from Eq.~\ref{eq:mag}. At the critical value $U = U_c$, the magnetization has a jump corresponding to the phase transition from $\ell=0$ to $\ell=1$.} 
\label{fig:mag_plot}
\end{figure}

We can further validate the jump in magnetization discussed in Section~\ref{sec:magnetization} by numerically computing the magnetization $\langle \hat{M}_z\rangle$ (Eq.~\ref{eq:mag_full}) utilizing the numerical solution for the wavefunction $f(p)$. An example is shown in Fig.~\ref{fig:mag_plot} for a specific value of $n$ within the WC phase. The abrupt jump in $\langle \hat{M}_z\rangle$ as a function of increasing $U$ is associated with the transition from $\ell = 0$ to $\ell = 1$.

So far, we have discussed the phase boundary of the WC state at zero temperature, for which the melting of the WC with increasing density arises from quantum fluctuations. But at finite temperatures, thermal fluctuations can also melt the WC state. At densities not too close to the critical density associated with quantum melting (at a given value of $U$), one can estimate the melting temperature by setting the Lindemann ratio to be $\eta=\sqrt{\left\langle r^{2}\right\rangle _{\text{thermal}}}/a$, where the amplitude of classical fluctuations is estimated using the equipartition theorem: $k\left\langle r^{2}\right\rangle _{\text{thermal}} = k_B T$. The maximum melting temperature can be estimated by using the value of $k$ associated with the largest density $n_c$ of the WC state (here, $n_c \approx 0.0012$, see Fig.~\ref{fig: PD}) and setting $\eta = \eta_c \approx 0.23$. This procedure gives a maximum melting temperature on the order of $\sim 10$\,K, with the melting temperature decreasing proportional to $n^{1/2}$ as the density is reduced.

\section{Wigner crystal phase with trigonal warping}
\label{sec:trigonal}

In the previous section, we derived the properties of the WC state -- including the critical density associated with melting and the critical displacement field associated with orbital magnetization -- using a description that neglects trigonal warping. While this assumption renders the problem more analytically tractable, at low electron density, the trigonal warping scale in BBG can easily become larger than the electron kinetic energy so that the electron system breaks up into small ``pockets'' of carriers at each valley \cite{zhou2022isospin, holleis2023ising, seiler2022quantum, lin2023spontaneous, Dong2023Isospin}. In this section, we consider how the results from the previous section are modified by trigonal warping.

We first study the phase diagram of the WC state and show that the structure of central and side pockets in the dispersion relation leads to an unusual doubly-reentrant behavior of the WC/FL phase boundary. We also show that trigonal warping precludes the possibility of the state with spontaneous orbital polarization ($\ell = 1$) presented in the previous section and instead leads to a state with nontrivial mini-valley ordering in the Wigner lattice \cite{calvera2022pseudo}.

Throughout this section, we are able to ignore the effects of interband dielectric screening discussed in Sec.~\ref{sec:dielectric}, since even at $U=0$ the trigonal warping leads to a linear rather than a parabolic dispersion near the band touching point, and in this case dielectric screening produces only a weak renormalization of the effective dielectric constant \cite{Ando_Screening_2006, Gorbar2002Magnetic, Joy2022Wigner}.

\subsection{Phase diagram of the WC and doubly-reentrant melting}

We can estimate the boundary between the WC and FL phases using the HO description, as in the previous section. We first consider the case of a relatively low displacement field, $U < t$. In real units, this inequality corresponds to $U \lesssim 50$\,meV, which for a boron nitride substrate translates to a displacement field smaller than $\approx 0.5$\,V/nm \cite{zhou2022isospin}.

\begin{figure}[htb]
\centering
\includegraphics[width=1.0 \columnwidth]{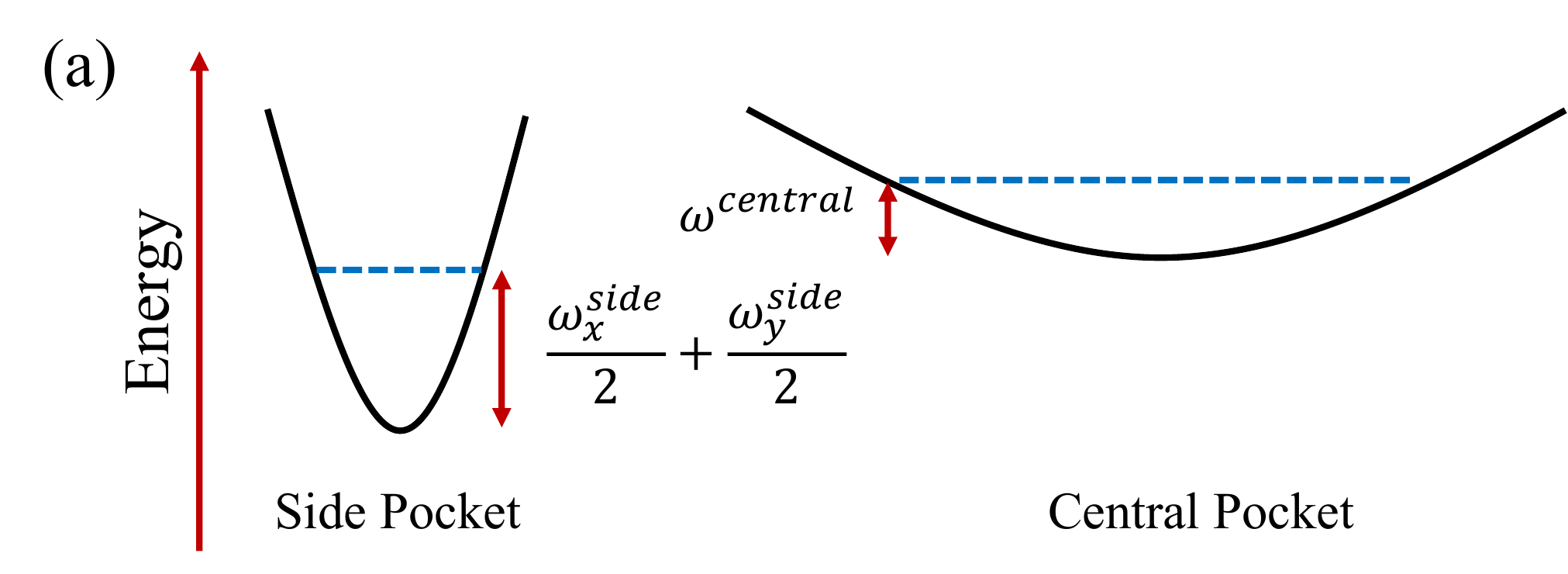}
\hspace{15 cm}
\bigskip
\includegraphics[width=1.0 \columnwidth]{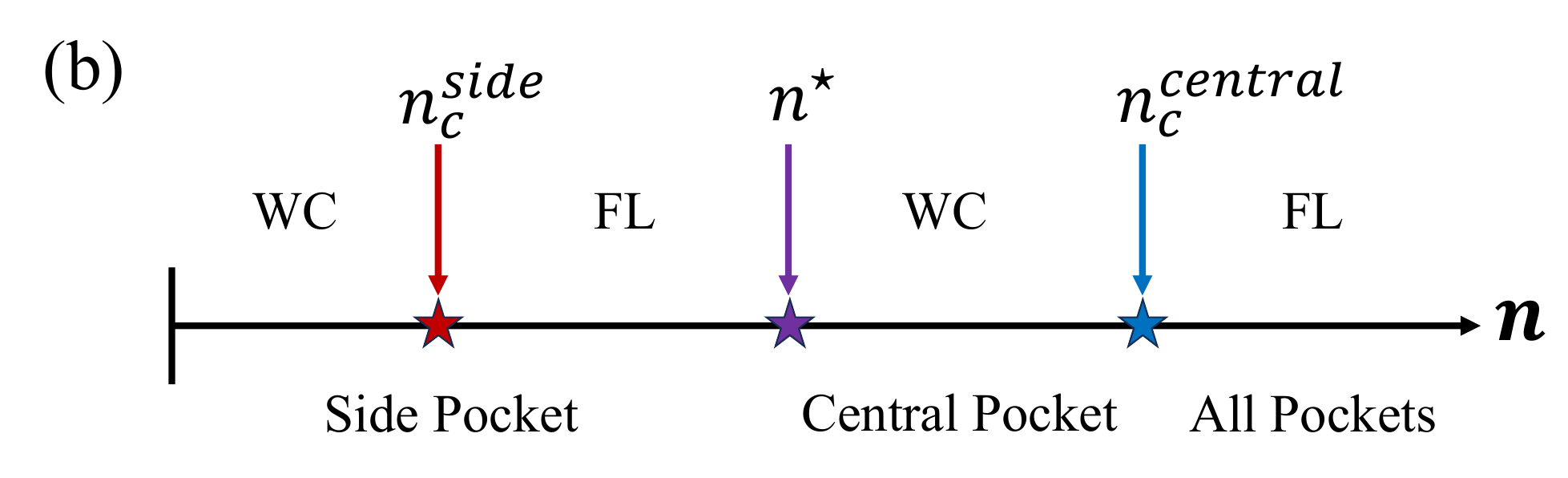}
\caption{(a) Schematic depiction of the side and central pocket dispersions and the corresponding energy of the HO ground state in each. In the situation depicted here, the electron resides in the side pocket despite its lighter mass. (b) Schematic line cut of the phase diagram at a fixed $U < t$. At very low density, Wigner crystallization occurs in the side pockets due to their lower band edge energy. As the density $n$ increases, the WC undergoes quantum melting at $n_{c}^\text{side}$, transitioning into a putative Fermi liquid phase in the side pockets. However, beyond a critical density $n^\star$, it becomes energetically favorable for crystallization to take place in the higher-energy central pocket. This central pocket WC eventually undergoes melting as the density exceeds $n_{c}^{central}$.} 
\label{fig: Reentrant}
\end{figure}

When $U<t$, the low energy conduction band comprises three side pockets and a central pocket having a dispersion given by
\begin{equation}
\varepsilon_{\text{central}}\left(\vec{p}\right)\simeq\frac{U}{2}+\frac{\left(t^{2}-U^{2}\right)}{U}\left(p_{x}^{2}+p_{y}^{2}\right).
\label{eq:central_dipersion}
\end{equation}
Notice that when $U$ exceeds $t$, the central pocket disappears and transforms into a dispersion maximum. 

The side pockets, meanwhile, have an anisotropic dispersion, which in the limit $U \ll t \ll 1$ can be approximated by
\begin{equation}
\varepsilon_{\text{side}}\left(\vec{p}\right)\simeq\frac{U}{2}-t^{2}U + \frac{t^{2}}{U} \tilde{p}_{x}^{2} + \frac{9t^{2}}{U}\tilde{p}_{y}^{2}.
\end{equation}
Here, $\tilde{p}_y$ and $\tilde{p}_y$ represent the components of momentum relative to the minimum of the pocket, with $\tilde{p}_x$ being in the direction directly away from the K point and $\tilde{p}_y$ being in the perpendicular direction.

Notice that the central pocket is slightly higher in energy than the side pockets but has a heavier mass (see Fig.~\ref{fig: Reentrant}a). Thus, whether the electron wavefunction lives primarily in the central or side pockets depends on the strength of the confining potential and, therefore, on the electron density. When the electron density is low, each electron in the WC resides primarily in the lower-energy side pockets. But when the electron density is high, the wavefunction has most of its weight in the heavier central pocket, which provides a lower confinement energy $\omega$. One can estimate the critical density $n^\star$ associated with the crossover from the side pockets to the central pocket  by equating the energies of the corresponding HO ground states. In the limit $U \ll t \ll 1$, this procedure gives
\be
n^{\star}\simeq0.023U^{2}.
\ee
At $n \ll n^{\star}$, the electron wavefunction resides primarily in the side pockets, while at $n \gg n^\star$, the electron wavefunction resides primarily in the central pocket. 

One can also estimate the critical densities associated with WC melting (via the Lindemann criterion) for both the central and side pockets. These give
\begin{equation}
n_c^{\text{central}} = (1.32\times10^{-5}) \frac{U^2}{(t^2-U^2)^2}.
\label{eq: nc_central}
\end{equation}
and
\begin{equation}
n_{c}^{\text{side}}\simeq0.004U^{2},
\end{equation}
respectively.

\begin{figure}[htb]
\centering
\includegraphics[width=1.0 \columnwidth]{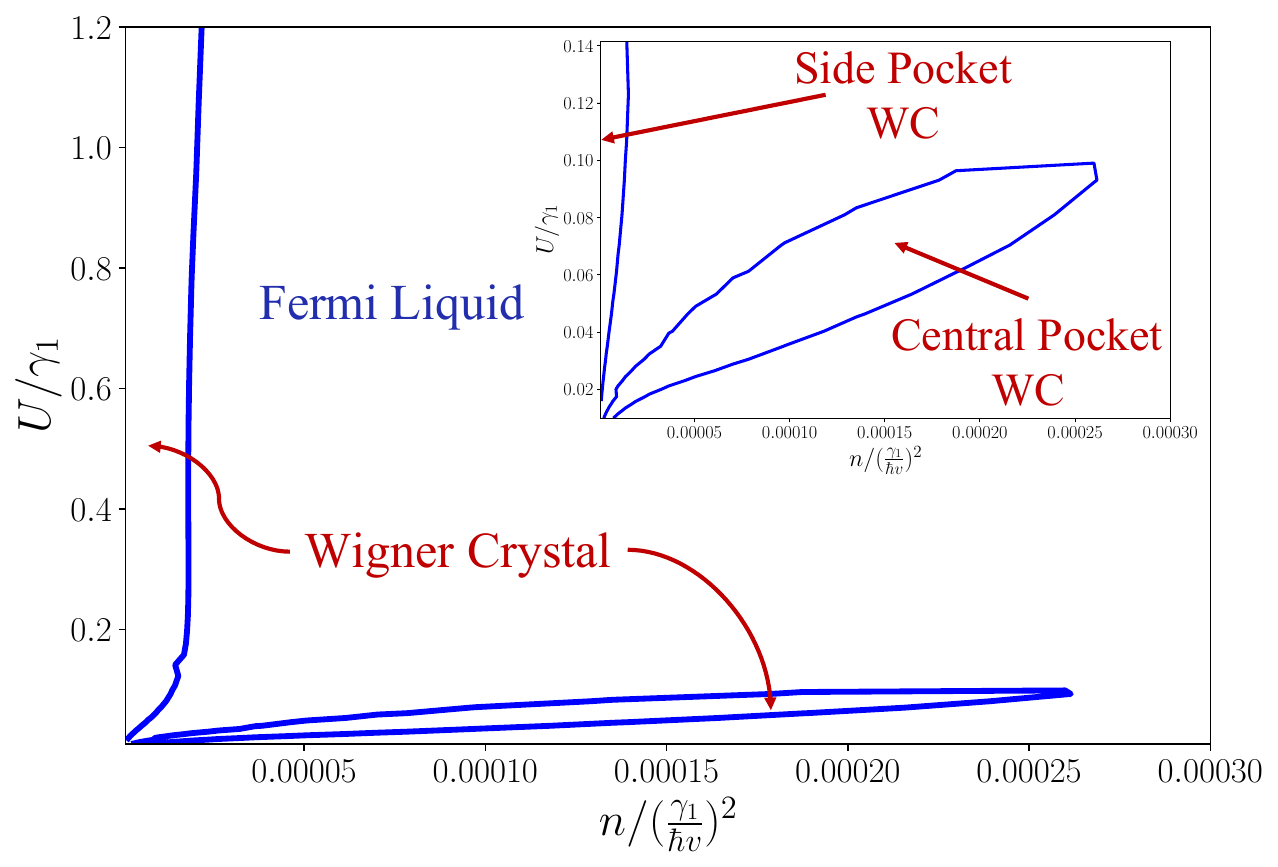}
\caption{The phase diagram of the low-density electron gas in BBG in the space of electron density $(n)$ and displacement field $(U)$ in the presence of trigonal warping. The blue line represents the phase boundary between Wigner crystal (WC) and Fermi Liquid (FL) and is calculated using setting Lindemann ratio $\eta= \eta_c = 0.23$. Inset shows that at small $U$, there is a doubly reentrant Wigner crystallization transition.} 
\label{fig: PD_twarp}
\end{figure}

Notice, however, that $n_{c}^{\text{side}} < n^\star < n_{c}^{\text{central}}$. This hierarchy suggests a scenario we refer to as ``doubly-reentrant melting'' (see Fig.~\ref{fig: Reentrant}b): as the density $n$ is increased from zero, the WC (in the side pocket) first melts at $n \approx n_{c}^{\text{side}}$, then the electron system transitions to the central pocket and freezes at $n \approx n^\star$, and then melts again at $n \approx n_{c}^{\text{central}}$. We confirm this scenario with a numerical solution in Fig.~\ref{fig: PD_twarp} (described below).

We note that Eq.~\ref{eq: nc_central} describes a critical density $n_c^{\text{central}}$ for the central pocket that diverges when $U$ approaches $t$ due to the diverging band mass of the central pocket. However, when the density $n$ is sufficiently large, the width $\sigma$ of the associated wave packet in momentum space becomes sufficiently large that one can no longer use a simple parabolic band approximation to describe the central pocket. Instead, when $\sigma$ becomes of the order of the distance between the central and side pockets ($\sim t$), the electron wavefunction leaks into the side pockets, and the WC state is eliminated. This effect truncates the divergence of $n_c^{\text{central}}$ and sets the maximum density $n_\textrm{max}$ associated with the WC state. Setting $\sigma \sim t$ and using the Lindemann criterion gives $n_\textrm{max} \sim \eta_c^2 t^2$, which is consistent with the maximum density of the WC state calculated numerically and shown in Fig.~\ref{fig: PD_twarp}.  The corresponding maximum melting temperature for the WC state, estimated using the procedure described in Sec.~\ref{sec:MHnumerics}, is of order $\sim 1$\,K.

On the other hand, in the limit of large $U$, where Wigner crystallization occurs only in the side pockets, we can estimate the melting density $n_c$ using the HO approximation for a single pocket. To zeroth order in $t$ (where the trigonal warping is ignored), $n_c$ remains unchanged from Eq.~\ref{eq: nclargeU}. Taking into account the correction arising from a large but finite mass along the $\tilde{p}_y$ direction, the modified critical density $n_c$ at large $U$ is given by
\begin{equation}
n_c \approx 2.1\times10^{-4}\left(1-6\sqrt{2}\sqrt{\frac{t}{U}}\right).
\end{equation}
We discuss the density range associated with the WC state in real units in Sec.~\ref{sec:conclusion}. 

It is worth noting that a qualitatively similar phase diagram to that of Fig.~\ref{fig: PD_twarp}, including a doubly-reentrant WC-like insulating phase, has been observed experimentally in rhombohedral pentalayer graphene \cite{Han2025signatures, han2026evidence}.
We note also that, in reality, as one approaches the WC-FL transition, an anisotropic lattice emerges, in addition to anisotropic electron wavepackets \cite{sammon2017electron}. This anisotropy has the effect of stabilizing the WC phase to somewhat higher concentrations $n_c$ than what is shown in Fig.~\ref{fig: PD_twarp}.

\subsection{Magnetization and orbital ordering}

In Sec.~\ref{sec:MHBC}, we showed that, for the case where there is no trigonal warping, Berry curvature induces a jump in the magnetization as a function of increasing $U$. This jump is associated with a transition from zero to finite angular momentum of the electron state at each Wigner lattice site. The presence of trigonal warping removes the rotational symmetry of the dispersion relation so that angular momentum is no longer a good quantum number. Nonetheless, within the HO description, there is still a jump in magnetization as a function of $U$, which can be understood as follows.

At sufficiently large $U > t$, only the side pockets are occupied by the electron wavefunction, and in general, the electron wavefunction resides in a superposition of these three pockets. Within the HO description, where the confining potential is rotationally symmetric and in the absence of Berry curvature, the ground state represents a symmetric superposition of the three pockets. The first excited state, in this case, is doubly degenerate and corresponds to a state in which the phase of the wavefunction (in momentum space) winds as a function of the angle $\phi$ by $2 \pi$ in either the clockwise or counterclockwise direction. These two excited states with nontrivial winding are associated with a nonzero orbital magnetization of the electron. In the presence of nonzero Berry curvature, the ordering of the ground state and first excited state is inverted if the Berry flux through the interior of the three pockets is greater than $\pi$, such that the state with nonzero orbital magnetization aligned with the Berry flux becomes the lowest energy state. This result can be viewed as the momentum space analog of the problem of an electron hopping among sites on an equilateral triangle that is pierced by magnetic flux.  A numerical calculation of the corresponding value of $U$ associated with the winding transition gives $U_c \approx 1.05$. A detailed quantitative discussion of the magnetization transition within the HO description is presented in Appendix~\ref{sec: winding}.

\begin{figure}[htb]
\centering
\includegraphics[width=1.0 \columnwidth]{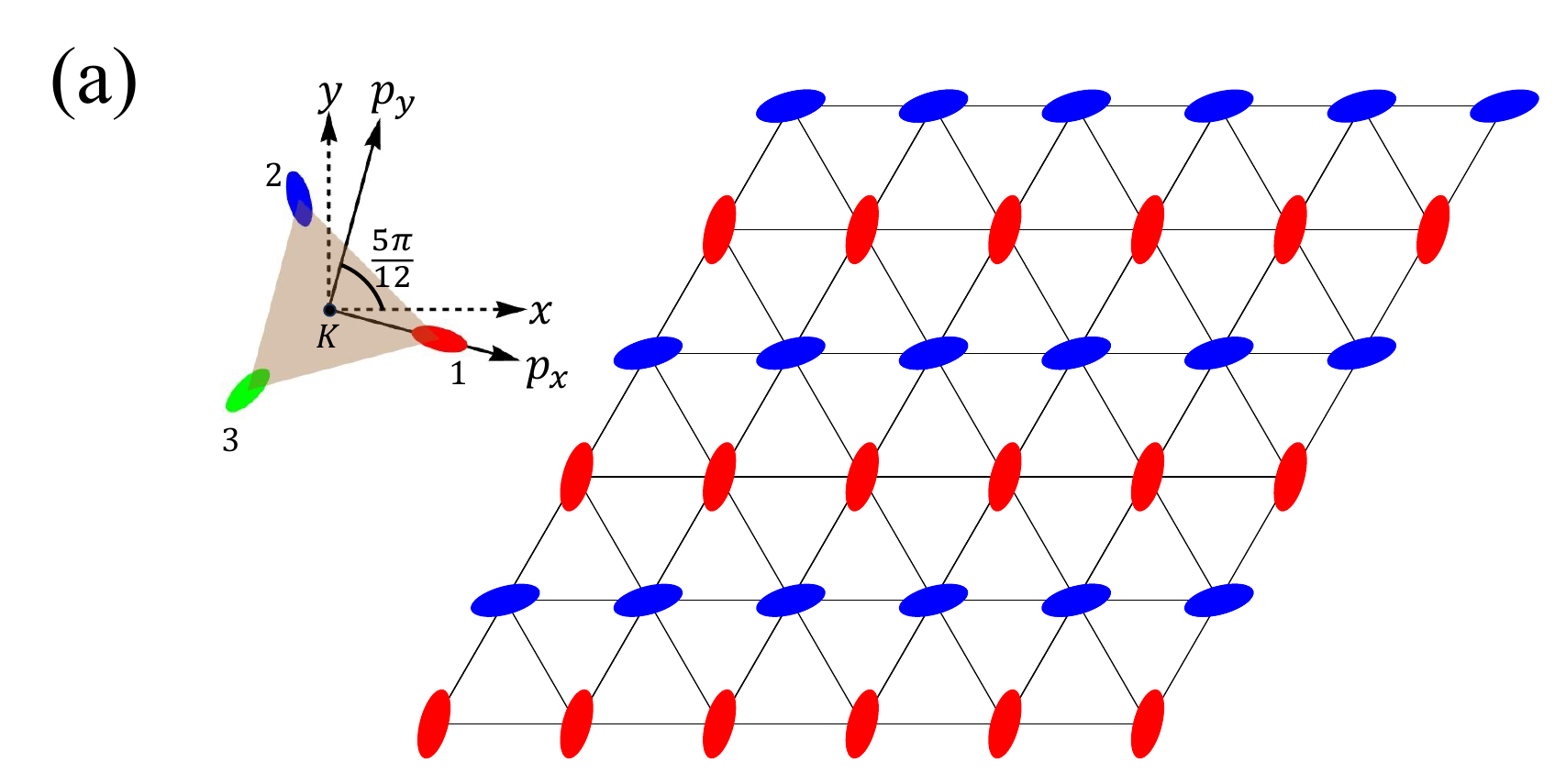}
\hspace{15 cm}
\bigskip
\includegraphics[width=1.0 \columnwidth]{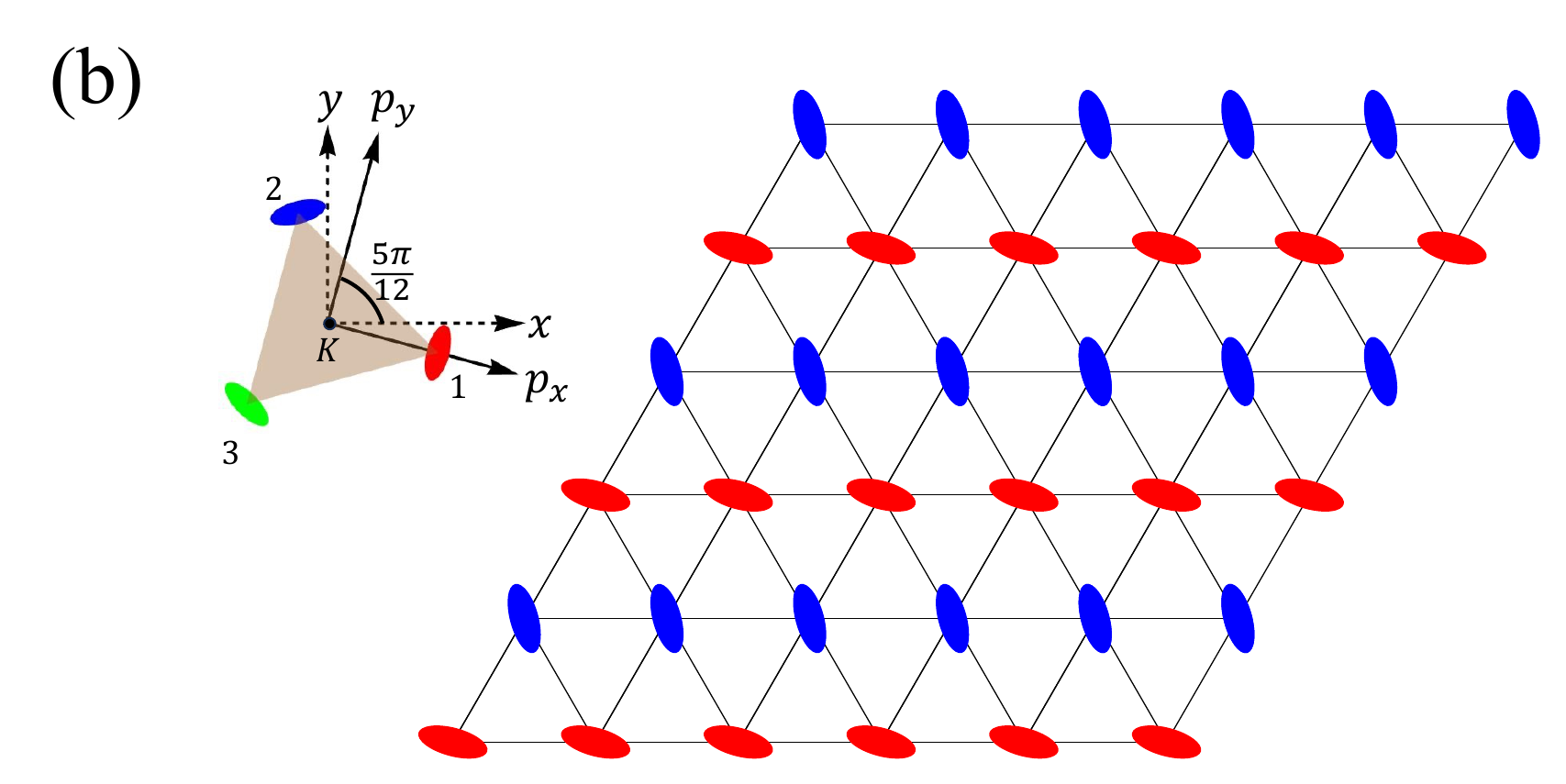}
\caption{Illustration of the pattern of minivalley polarization in the ground state of the WC (following Ref.~\cite{calvera2022pseudo}). This pattern involves alternating stripes of two minivalleys (labeled ``$1$''/``$2$'' and colored red/blue, respectively). Here we depict the WC in real space (main figures) as well as the minivalleys/pockets in momentum space (insets). 
(a) When $U <  U_\textrm{iso}$, the pockets/minivalleys are elongated toward the $K$ point. (b) When $U >  U_\textrm{iso}$, the pockets/minivalleys are elongated in the direction perpendicular to the $K$ point.}
\label{fig: Stripe_MV}
\end{figure}

In practice, however, the energy scale $\tau$ associated with splitting between the winding and non-winding energy states is exponentially small in the inverse density: $\log \tau \propto - (p_0/\sigma)^2 \propto -1/n^{3/4}$. Consequently, the actual minivalley ordering of electrons in the WC is dominated by effects that are beyond the HO description. In particular, the correction to the WC energy arising from spatial anisotropy of the electron wavefunction is of order $n^{3/4}$ (commonly written as $1/r_s^{3/2}$, where $r_s$ is the usual interaction parameter). The optimal minivalley ordering that minimizes the WC energy is a nontrivial problem, but for the case of three pockets related by $C_3$ symmetry, this problem was, in fact, studied recently by Calvera et.~al.\ in Ref.~\cite{calvera2022pseudo}. These authors found that the minimal energy state of the WC is for each electron to be polarized into one of two minivalleys, with an alternating stripe pattern of minivalley polarization on the Wigner lattice. This arrangement is depicted schematically in Fig.~\ref{fig: Stripe_MV}.

Thus, in our problem, the regime of the ``side pocket WC'' presumably corresponds to the nontrivial minivalley-ordered state depicted in Fig.~\ref{fig: Stripe_MV}. The regime of larger $n$ and ${U < t}$, for which only the central pocket is occupied, is trivially ordered.

It is interesting to note that the anisotropy of the three side pockets changes sign as a function of $U$. When $U$ is smaller than a specific value $U_\textrm{iso} = 9t/2^{3/2}$, the pockets are elongated  toward the $K$ point, while at $U$ larger than this value the pockets are elongated in the direction perpendicular to the $K$ point (see Fig.~\ref{fig:DispersionTwarp}). This geometric change implies that when $U$ is close to $U_\textrm{iso}$ the pockets become isotropic, and the system is indifferent with respect to minivalley ordering. At this point, one can expect the WC to acquire a particularly large entropy associated with the minivalley degree of freedom. In principle, this increase in entropy of the WC state should lead to an increased melting temperature of the WC state when $U$ is close to $U_\textrm{iso}$, as in the Pomeranchuk effect \cite{pomeranchuk1950on, saito2021isospin}. 

\subsection{Numerical results}

In order to numerically calculate the phase diagram with trigonal warping, it is necessary to solve the two-dimensional Schrödinger equation (see Eq.~\ref{eq:H_eff_2}) in momentum space. (Unlike in the case without trigonal warping, here we do not have rotational symmetry and therefore cannot reduce the problem to an effective 1D Schrödinger equation.) We include the effects of Berry curvature by defining a Berry connection relative to an arbitrarily chosen point in momentum space (we choose $\vec{p} = 0$) via an integral of the Berry curvature. The details of this method are described in Appendix \ref{sec: NumBConn}.

We diagonalize the Schrödinger equation on a discrete triangular grid in momentum space (with a size of approximately $200 \times 200$ points). From the corresponding lowest-energy wavefunction, we calculate the Lindemann ratio, which allows us to estimate the phase diagram associated with the melting of the WC state. The result is shown in Fig.~\ref{fig: PD_twarp}. The double re-entrance of the WC phase as a function of density, associated with electrons crystallizing first in the side pockets and then the central pocket, can be seen clearly at small $U$. 

\section{Conclusions and comments on experiments}
\label{sec:conclusion}

In summary, here we have considered the WC state in BBG, providing a discussion of its ground state ordering and an estimation of its phase boundary in the space of density and displacement field. One of our more remarkable conclusions is that if one ignores trigonal warping, then the radially-symmetric low energy band structure enables a magnetization transition of the WC state as a function of the increasing displacement field, driven by the coupling between the Berry flux $\Phi$ and the orbital angular momentum $\ell$.
Unfortunately, trigonal warping drastically lowers the energy scale associated with this coupling since it splits the rotationally-symmetric ``Mexican hat'' band structure into three discrete mini-valleys. In principle, an electron in a radially symmetric confining potential still undergoes a magnetization transition as a function of the displacement field, even when trigonal warping is present, associated with a winding transition in the momentum space wavefunction. Such a transition may be relevant for quantum dots in BBG \cite{ge_visualization_2020}.  In the WC, however, the small energy scale associated with the coupling between $\Phi$ and $\ell$ is overwhelmed by the non-radially-symmetric component of the confining potential arising from the Wigner lattice. Instead, we predict (using the results of Ref.~\cite{calvera2022pseudo}) that the ground state of the WC is associated with a nontrivial patterning of the minivalley order in space (Fig.~\ref{fig: Stripe_MV}). It is worth noting that a very recent theoretical study \cite{Aguilar2026full} using Hartree-Fock calculations has predicted a qualitatively similar phase diagram to our Fig.~\ref{fig: PD_twarp}, including multiple reentrances of the WC state, persisting to much higher density (with different regions of the WC being associated with different patterns of spin and valley polarization). However, Hartree-Fock is known to over-predict the stability of the WC phase by as much as two orders of magnitude in electron density \cite{azadi2024quantum, bernu2011hartree, Aguilar2026full}.

In general, the existence of a WC state can be inferred experimentally by a set of distinctive features in transport and thermodynamic measurements. WCs are insulators in terms of their temperature-dependent conductivity \cite{shklovskii_coulomb_2004}, with the I-V curve at low temperature exhibiting a sharp ``pinning voltage" that is often hysteretic \cite{yoon_wigner_1999, knighton_evidence_2018, falson_competing_2022}. WCs also exhibit ``negative compressibility" in capacitance or penetration field measurements, which is a hallmark of strong positional correlations between electrons (see, e.g., Refs.~\cite{bello_density_1981, kravchenko1990evidence, eisenstein1992negative, eisenstein1994compressibility, shapira1996thermodynamics, Skinner2010anomalously, Li2011very, skinner2013giant}). The experiments of Ref.~\cite{zhou2022isospin} observed negative compressibility emerging at low temperature and high displacement field but not coexisting with an insulating temperature dependence. Ref.~\cite{seiler2022quantum} reports a state with insulating-like temperature dependence that emerges within a window of relatively high displacement field and low densities $n \approx 1.5 \times 10^{11}$\,cm$^{-2}$, which the authors characterize as being consistent with a WC. Our calculations here, on the other hand, suggest that the maximum density associated with the WC state (occurring at displacement field $D \approx 0.7$\,V/nm) is only of order $1.2 \times 10^{10}$\,cm$^{-2}$ (see Fig.~\ref{fig: PD_twarp}). Such low densities are typically difficult to probe experimentally since disorder in the sample introduces an energy scale that randomly modulates the electron density. Experiments on BBG in a displacement field all observe a strongly insulating state with positive compressibility at sufficiently low density and a large displacement field, which presumably corresponds to this disorder-dominated situation. At $D \approx 0.7$\,V/nm the strongly insulating state occupies a range of density $2$ -- $6 \times 10^{10}$\,cm$^{-2}$, varying from one experiment to another \cite{zhou2022isospin, seiler2022quantum, de2022cascade, lin2023spontaneous}. Thus the WC regime that we predict seems to be only barely outside the range of current experiments. It will be interesting to see whether future experiments can confirm the re-entrant melting of the WC state in BBG. In principle, one can also infer the existence of a WC optically, using features of the exciton absorption spectrum \cite{smolenski_signatures_2021, zhou_bilayer_2021}. Unfortunately, such experiments are technically difficult in BBG due to the very small band gap.

Direct observation of the nontrivial pattern of minivalley order depicted in Fig.~\ref{fig: Stripe_MV} could, in principle, be accomplished by scanning microscopy. But a perhaps more salient feature would be to observe the associated Pomeranchuk effect that arises at $U = U_\textrm{iso} \approx 150$\,meV ($D \approx 1.5$\,V/nm) due to the large configurational entropy of the WC when the minivalleys become isotropic. Unfortunately, this effect appears only in the lower-density regime of the WC associated with crystallization in the side pockets, which by our (conservative) estimation, is limited to very low densities $n \lesssim 2 \times 10^9$\,cm$^{-2}$. 

Finally, although we have focused in this paper on BBG, it is worth emphasizing that in the last few years (since our preprint was first posted \cite{joy2023wignerv1}), rhombohedral-stacked graphene with a higher number $N$ of layers has emerged as an especially promising platform for realizing unconventional WC phases. In rhombohedral pentalayer graphene, for example, a WC-like insulating phase extends to densities as high as $5 \times 10^{11}$\,cm$^{-2}$ \cite{Han2025signatures, lu2024fractional, han2026evidence}, more than one hundred times larger than the value estimated above for BBG. This improved stability of the WC phase can be attributed to the flatter band structure associated with higher-$N$ rhombohedral graphene and to the reduced energy scale associated with trigonal warping. (These experiments have also observed a double reentrance of the WC phase, similar to what is shown in our Fig.~\ref{fig: PD_twarp} \cite{Han2025signatures, han2026evidence}.) The larger Berry curvature also makes these materials more ideal platforms for realizing Berry-phase-driven WC phases, since in $N$-layer rhombohedral graphene there is $N\pi$ Berry flux in the vicinity of the $K$ and $K'$ points, rather than just $2\pi$. The larger Berry flux can also facilitate the formation of ``quantum anomalous Hall crystal" phases, in which a WC-like bulk is accompanied by a dissipationless edge state and quantized Hall conductance~\cite{tesanovic1989hall, Dong2024anomalous, soejima2024anomalous, dong2024stability, tan2024parent, zeng2024sublattice, tan2025variational, zhou2025new, soejima2025lambda, miao2026various}.

It is important to note that the orbitally-polarized state we discuss here is distinct from the quantum anomalous Hall crystal, and indeed the ``Einstein phonon" approximation we employ is incapable of reproducing phases with nontrivial Chern number. In this sense one can think of the phase diagram of the WC state as generally being characterized by two distinct quantities: the orbital angular momentum $\ell$ and the Chern number $C$. A full phase diagram of the WC remains to be explored, although some preliminary work has been presented in Ref.~\cite{dong2024stability, Dong2024anomalous, soejima2025lambda}. (Within the model studied in Ref.~\cite{soejima2025lambda}, the analog of our $\ell \neq 0$ phase is dubbed the ``halo WC" phase).

\acknowledgments
We gratefully acknowledge Leonid Levitov for his collaboration on the companion paper, Ref.~\cite{joy2025chiral}. The authors are grateful also to Zachariah Addison, Zhiyu Dong, Liang Fu, Aaron Hui, Kyle Kawagoe, Steve Kivelson, T.~Senthil, Tomohiro Soejima, and Jairo Velasco for useful discussions. This work was supported by the NSF under Grant No.~DMR-2045742. 
\bibliography{WC+OM}
\newpage
\onecolumngrid
\appendix
\section{Derivation of the effective harmonic oscillator Hamiltonian in a band with Berry curvature}
\label{sec:HO_BC}

In this section, we derive the effective HO Hamiltonian in momentum space when the electrons originate from a band with non-zero Berry curvature \cite{Price2014Quantum, Berceanu2016Momentum, Karplus1954Hall, Lapa2019Semiclassical}. The single particle Hamiltonian is given by $H=H_{0}+U\left(r\right)$, where $H_{0}$ comes from the underlying crystal lattice and $U\left(r\right)=kr^{2}/2$ is the harmonic confinement potential. The Bloch eigenstates of $H_{0}$ are denoted by $\left|n,p\right\rangle$  ($n$ and $\vec{p}$ denote the band index and quasi-momentum), which satisfies $H_{0}\exp\left(\dot{\iota}\vec{p}\cdot\vec{r}\right)\left|n,p\right\rangle =E_{n}\left(\vec{p}\right)\exp\left(\dot{\iota}\vec{p}\cdot\vec{r}\right)\left|n,p\right\rangle$, where $E_{n}\left(\vec{p}\right)$ is the band dispersion of the $n^{\text{th}}$ band. Any eigenstates of $H$ can be expanded in the eigenbasis of $H_{0}\left(\vec{p}\right)$ as
\begin{equation}
\left|\psi\right\rangle =\sum_{n}\sum_{\vec{p}}\psi_{n}\left(\vec{p}\right)\exp\left(\dot{\iota}\vec{p}\cdot\vec{r}\right)\left|n,p\right\rangle, 
\end{equation}
where $\psi_{n,\vec{p}}$ denotes the expansion coefficients. The eigenvalue equation for the Hamiltonian $H$ can be written as
\begin{equation}
\sum_{n}\sum_{\vec{p}}E_{n}\left(\vec{p}\right)\psi_{n}\left(\vec{p}\right)\exp\left(\dot{\iota}\vec{p}\cdot\vec{r}\right)\left|n,p\right\rangle +U\left(r\right)\sum_{n}\sum_{\vec{p}}\psi_{n}\left(\vec{p}\right)\exp\left(\dot{\iota}\vec{p}\cdot\vec{r}\right)\left|n,p\right\rangle =\epsilon\sum_{n}\sum_{\vec{p}}\psi_{n}\left(\vec{p}\right)\exp\left(\dot{\iota}\vec{p}\cdot\vec{r}\right)\left|n,p\right\rangle. 
\end{equation}
For brevity, let us denote $\left|\chi_{n,\vec{p}}\right\rangle =\exp\left(\dot{\iota}\vec{p}\cdot\vec{r}\right)\left|n,p\right\rangle$. Taking an inner product with $\left\langle \chi_{m,\vec{q}}\right|$, the above equation yields
\begin{equation}
E_{m}\left(\vec{q}\right)\psi_{m}\left(\vec{q}\right)+\sum_{n}\sum_{\vec{p}}\left\langle \chi_{m,\vec{q}}\right|U\left(r\right)\left|\chi_{n,\vec{p}}\right\rangle \psi_{n}\left(\vec{p}\right)=\epsilon\psi_{m}\left(\vec{q}\right).
\end{equation}
We used the orthonormality condition of the Bloch eigenstates $\left\langle \chi_{m,\vec{q}}\right.\left|\chi_{n,\vec{p}}\right\rangle =\delta\left(\vec{p}-\vec{q}\right)\delta_{m,n}$. We are left with the evaluation of $\left\langle \chi_{m,\vec{q}}\right|U\left(r\right)\left|\chi_{n,\vec{p}}\right\rangle$. Let us first evaluate $\left\langle \chi_{m,\vec{q}}\right|\widehat{\vec{r}}\left|\chi_{n,\vec{p}}\right\rangle$. Using the identity $\int d\vec{r}\left|\vec{r}\right\rangle \left\langle \vec{r}\right|=\mathbb{I}$ and $\left\langle \vec{r}\right.\left|\chi_{n,\vec{p}}\right\rangle =\exp\left(\dot{\iota}\vec{p}\cdot\vec{r}\right)w_{n,\vec{p}}\left(\vec{r}\right)$, where $w_{n,\vec{p}}\left(\vec{r}\right)\equiv\left\langle \vec{r}\right.\left|n,\vec{p}\right\rangle$ we obtain
\begin{equation}
\left\langle \chi_{m,\vec{q}}\right|\widehat{\vec{r}}\left|\chi_{n,\vec{p}}\right\rangle=\int d\vec{r}\exp\left(-\dot{\iota}\vec{q}\cdot\vec{r}\right)w_{m,\vec{q}}^{*}\left(\vec{r}\right)\vec{r}\exp\left(\dot{\iota}\vec{p}\cdot\vec{r}\right)w_{n,\vec{p}}\left(\vec{r}\right).
\end{equation}
Using the following relation, $\vec{r}\exp\left(\dot{\iota}\vec{p}\cdot\vec{r}\right)=-\dot{\iota}\vec{\nabla}_{\vec{p}}\exp\left(\dot{\iota}\vec{p}\cdot\vec{r}\right)$, the integrand in the previous equation can be rewritten
\begin{equation}
\left\langle \chi_{m,\vec{q}}\right|\widehat{\vec{r}}\left|\chi_{n,\vec{p}}\right\rangle=\dot{\iota}\vec{\nabla}_{\vec{p}}\left(\int d\vec{r}\exp\left(\dot{\iota}\left(\vec{p}-\vec{q}\right)\cdot\vec{r}\right)w_{m,\vec{q}}^{*}\left(\vec{r}\right)w_{n,\vec{p}}\left(\vec{r}\right)\right)+\dot{\iota}\int d\vec{r}\exp\left(\dot{\iota}\left(\vec{p}-\vec{q}\right)\cdot\vec{r}\right)w_{m,\vec{q}}^{*}\vec{\nabla}_{\vec{p}}w_{n,\vec{p}}\left(\vec{r}\right).
\label{eq:A5}
\end{equation}
The first term in the above equation can be evaluated to be
\begin{equation}
-\dot{\iota}\vec{\nabla}_{\vec{p}}\left(\int d\vec{r}\exp\left(\dot{\iota}\left(\vec{p}-\vec{q}\right)\cdot\vec{r}\right)w_{m,\vec{q}}^{*}\left(\vec{r}\right)w_{n,\vec{p}}\left(\vec{r}\right)\right)=\dot{\iota}\delta\left(\vec{q}-\vec{p}\right)\vec{\nabla}_{\vec{p}}\delta_{m,n}.
\end{equation}
The second integral in Eq.~\ref{eq:A5} vanishes unless $\vec{p}=\vec{q}$, because $w_{n,\vec{p}}\left(\vec{r}\right)$ is a periodic function, which gives
\begin{equation}
\dot{\iota}\int d\vec{r}\exp\left(\dot{\iota}\left(\vec{p}-\vec{q}\right)\cdot\vec{r}\right)w_{m,\vec{q}}^{*}\vec{\nabla}_{\vec{p}}w_{n,\vec{p}}\left(\vec{r}\right)=\dot{\iota}\delta\left(\vec{q}-\vec{p}\right)\left\langle m,\vec{q}\right|\vec{\nabla}_{\vec{q}}\left|n,\vec{q}\right\rangle.
\end{equation}
Recognizing the (non-abelian) Berry connection $A_{m,n}\left(\vec{q}\right)=\dot{\iota}\left\langle m,\vec{q}\right|\vec{\nabla}_{\vec{q}}\left|n,\vec{q}\right\rangle$ , we arrive at
\begin{equation}
\left\langle \chi_{m,\vec{q}}\right|\widehat{\vec{r}}\left|\chi_{n,\vec{p}}\right\rangle =\delta\left(\vec{q}-\vec{p}\right)\left(\dot{\iota}\delta_{m,n}\vec{\nabla}_{\vec{p}}+A_{m,n}\left(\vec{p}\right)\right).
\end{equation}
Similarly, inserting the completeness identity twice,  $\widehat{\vec{r}}^2$ can be shown to be
\begin{equation}
\sum_{n}\sum_{\vec{p}}\left\langle \chi_{m,\vec{q}}\right|\widehat{\vec{r}}^2\left|\chi_{n,\vec{p}}\right\rangle \psi_{n}\left(\vec{p}\right)=\sum_{n,n'}\left(\dot{\iota}\delta_{m,n'}\vec{\nabla}_{\vec{q}}+A_{m,n'}\left(\vec{q}\right)\right)\cdot\left(\dot{\iota}\delta_{n',n}\vec{\nabla}_{\vec{q}}+A_{n',n}\left(\vec{q}\right)\right)\psi_{n}\left(\vec{q}\right),
\end{equation}
leading to the following SE
\begin{equation}
E_{m}\left(\vec{q}\right)\psi_{m}\left(\vec{q}\right)+\frac{k}{2}\sum_{n,n'}\left(\dot{\iota}\delta_{m,n'}\vec{\nabla}_{\vec{q}}+A_{m,n'}\left(\vec{q}\right)\right)\cdot\left(\dot{\iota}\delta_{n',n}\vec{\nabla}_{\vec{q}}+A_{n',n}\left(\vec{q}\right)\right)\psi_{n}\left(\vec{q}\right)=\epsilon\psi_{m}\left(\vec{q}\right).
\end{equation}
We have not made any assumptions in deriving the above equation. Now, let's assume that the relevant band index is $m$, and in the second term, only $\psi_{m}\left(\vec{q}\right)$ is non-negligible. This lead to
\begin{equation}
E_{m}\left(\vec{q}\right)\psi_{m}\left(\vec{q}\right)+\frac{k}{2}\sum_{n'}\left(\dot{\iota}\delta_{m,n'}\vec{\nabla}_{\vec{q}}+A_{m,n'}\left(\vec{q}\right)\right)\cdot\left(\dot{\iota}\delta_{n',m}\vec{\nabla}_{\vec{q}}+A_{n',m}\left(\vec{q}\right)\right)\psi_{m}\left(\vec{q}\right)\approx\epsilon\psi_{m}\left(\vec{q}\right).
\end{equation}
Separating $n'=m$ and $n'\neq m$ terms, we can write
\begin{align}
    \begin{split}
        E_{m}\left(\vec{q}\right)\psi_{m}\left(\vec{q}\right)+\frac{k}{2}\left(\dot{\iota}\vec{\nabla}_{\vec{q}}+A_{m,m}\left(\vec{q}\right)\right)^{2}\psi_{m}\left(\vec{q}\right)+\frac{k}{2}\sum_{n'\neq m}\left(A_{m,n'}\left(\vec{q}\right)\right)\cdot\left(A_{n',m}\left(\vec{q}\right)\right)\psi_{m}\left(\vec{q}\right)&\approx\epsilon\psi_{m}\left(\vec{q}\right),\\\implies\left[E_{m}\left(\vec{q}\right)+\frac{k}{2}\left(\dot{\iota}\vec{\nabla}_{\vec{q}}+A_{m,m}\left(\vec{q}\right)\right)^{2}+\frac{k}{2}\sum_{n'\neq m}\left|A_{m,n'}\left(\vec{q}\right)\right|^{2}\right]\psi_{m}\left(\vec{q}\right)&\approx\epsilon\psi_{m}\left(\vec{q}\right).
    \end{split}
\end{align}
We have finally arrived at the effective Hamiltonian
\begin{equation}
H_{eff}=E_{m}\left(\vec{q}\right)+\frac{k}{2}\left(\dot{\iota}\vec{\nabla}_{\vec{q}}+A_{m,m}\left(\vec{q}\right)\right)^{2}+\frac{k}{2}\sum_{n'\neq m}\left|A_{m,n'}\left(\vec{q}\right)\right|^{2}.
\end{equation}
The energy dispersion and Berry connection of the $m^{th}$ band act as a momentum space scalar and magnetic vector potential. There is another contribution to the momentum space scalar potential given by
\begin{equation}
E_{add}\left(\vec{q}\right)=\frac{k}{2}\sum_{n'\neq m}\left|A_{m,n'}\left(\vec{q}\right)\right|^{2}.
\end{equation}
This additional term is the same as the trace of the underlying quantum geometric tensor \cite{Classen2015Position}. Using the following representation of $A_{n,n'}$, we can recast the additional term into a gauge invariant form
\begin{equation}
A_{n',n}=\dot{\iota}\frac{\left\langle n',\vec{p}\right|\left(\vec{\nabla}_{\vec{p}}H_{0}\left(\vec{p}\right)\right)\left|n,p\right\rangle }{\left(E_{n}\left(\vec{p}\right)-E_{n'}\left(\vec{p}\right)\right)}.
\end{equation}
The additional potential can be rewritten as
\begin{align}
    \begin{split}
        E_{add}\left(\vec{q}\right)&=\frac{k}{2}\sum_{n'\neq m}\left|A_{m,n'}\left(\vec{q}\right)\right|^{2},\\&=\frac{k}{2}\sum_{n'\neq m}\frac{\left|\left\langle m,\vec{q}\right|\left(\vec{\nabla}_{\vec{p}}H_{0}\left(\vec{p}\right)\right)\left|n',q\right\rangle \right|^{2}}{\left(E_{n'}\left(\vec{q}\right)-E_{m}\left(\vec{q}\right)\right)^{2}}.
    \end{split}
\end{align}

This last term $E_{add}(\vec{q})$ can be neglected in the WC regime where the electron density $n$ is small, since $E_{add}$ scales linearly with $k \propto n^{3/2}$ while the terms we are focus on are proportional to $\omega\propto \sqrt{k}\propto n^{3/4}$ (see Eq.~\ref{eq:spectrum_wo}).

\section{Harmonic Oscillator with Mexican hat dispersion}
\label{sec: MHHO}

In this section, we solve the problem of a particle confined in a Harmonic oscillator potential with a Mexican hat (MH) dispersion in the absence of Berry curvature. To that end, let us consider the following Hamiltonian:
\begin{equation}
\hat{H}=\frac{1}{2m}\left(\left|\hat{\vec{p}}\right|-p_{0}\right)^{2}+\frac{1}{2}m\omega^{2}\hat{\vec{r}}^{2},
\end{equation}
where $\hat{\vec{p}}$ and $\hat{\vec{r}}$ are the momentum and position operators, $\omega$ is the angular frequency, and $m$ is the mass of the particle. Since the position operator assumes the form of a gradient operator in momentum space, $\hat{\vec{r}}=\dot{\iota}\hbar\vec{\nabla}{\vec{p}}$, this problem look like a particle in a MH potential in momentum space. The corresponding Schrödinger equation (SE) in momentum space is given by:
\begin{equation}
\left[-\frac{1}{2}\hbar^{2}m\omega^{2}\left(\frac{\partial^{2}}{\partial p^{2}}+\frac{1}{p}\frac{\partial}{\partial p}+\frac{1}{p^{2}}\frac{\partial^{2}}{\partial\phi^{2}}\right)+\frac{1}{2m}\left(p-p_{0}\right)^{2}\right]\psi\left(\vec{p}\right)=E\psi\left(\vec{p}\right).
\label{eq: SE_MS}
\end{equation}
Due to the rotational symmetry of the problem, the solution to the above SE is of the variable separable form, $\psi\left(\vec{p}\right)=f\left(p\right)e^{\dot{\iota}\ell\phi}$, and the allowed values of $\ell$ are integers. In the first step, we turn the above SE into an effective one-dimensional SE. In order to do that, we use the ansatz $f=g/\sqrt{p}$, which allows us to eliminate the first-order derivative term. Substituting this ansatz, the SE becomes:
\begin{equation}
-\frac{1}{2}\hbar^{2}m\omega^{2}\left[\frac{d^{2}g}{dp^{2}}+\frac{1}{p^{2}}\left(\frac{1}{4}-\ell^2\right)g\right] + \frac{1}{2m}\left(p-p_{0}\right)^{2}g=Eg.
\end{equation}
We can recast the effective radial (in momentum space) SE in a dimensionless form by dividing all the momentum scales by $\sigma\equiv\sqrt{\hbar m\omega}$, so that $u\equiv p/\sigma$ and $u_{0}\equiv p_{0}/\sigma$. We obtain:
\begin{equation}
-\left[\frac{d^{2}g}{du^{2}}+\frac{1}{u^{2}}\left(\frac{1}{4}-\ell^{2}\right)g\right]+\left(u-u_{0}\right)^{2}g=\epsilon g
\label{Eq: SESimple},
\end{equation}
where we have defined $\epsilon\equiv E\big/(\hbar\omega/2)$. In the limit where the MH is deep enough, $u_{0}>>1$ (the wavefunction is a thin ring in momentum space), we can find the eigenstates and eigenenergy perturbatively in $1/u_0$. Let us expand the above equation around $u_0$ by taking $x\equiv u-u_{0}$ and keeping terms up to $1/u_0^2$:
\begin{equation}
-\frac{d^{2}g}{dx^{2}}+x^{2}g=\tilde{\epsilon}g,
\label{eq:1D_HO}
\end{equation}
where we defined:
\begin{equation}
\tilde{\epsilon}=\epsilon+\frac{1}{4u_{0}^{2}}-\frac{\ell^{2}}{u_{0}^{2}}.
\end{equation}
The SE in Eq.~\ref{eq:1D_HO} is the equation for the one-dimensional Harmonic oscillator, and its eigenvalues are given by:
\begin{equation}
\tilde{\epsilon} = 2n+1, \quad n\in\mathbb{Z}^{+}.
\end{equation}
Going back to the original units, the energy eigenvalues are given by:
\begin{equation}
E_{n,\ell}=\frac{\hbar\omega}{2}\left(2n+1\right)+\frac{\hbar\omega\ell^{2}}{2u_{0}^{2}}-\frac{\hbar\omega}{8u_{0}^{2}}.
\end{equation}
This result has a nice physical interpretation. When $p\sim p_{0}$, the particle effectively experiences harmonic confinement in the radial direction and behaves dispersionless in the azimuthal direction. Therefore, the ground state energy of this system corresponds to a one-dimensional harmonic oscillator. The corresponding normalized wavefunction for $n=0$ principal quantum number is given by:
\begin{equation}
\psi_\ell\left(p,\phi\right)=\left(\frac{2\sqrt{\pi}}{\sigma}\right)^{1/2}\frac{\exp\left[-\frac{\left(p-p_{0}\right)^{2}}{2\sigma^{2}}\right]}{\sqrt{p}}\:\exp\left[\dot{\iota}\ell\phi\right].
\label{eq:psi_p}
\end{equation}
By Fourier transformation, we can calculate the real space wavefunction:
\begin{align}
    \begin{split}
     \tilde{\psi}_\ell\left(r,\theta\right)&=\frac{1}{\left(2\pi\right)^{2}}\int d^{2}\vec{p} \, e^{\dot{\iota}\vec{p}.\vec{r}}\psi\left(\vec{p}\right),\\&=\left(\frac{2\sqrt{\pi}}{\sigma}\right)^{1/2}\frac{1}{\left(2\pi\right)^{2}}\int_{0}^{2\pi}d\phi\int_{0}^{\infty}pdp\frac{\exp\left[-\frac{\left(p-p_{0}\right)^{2}}{2\sigma^{2}}\right]}{\sqrt{p}}\:e^{\dot{\iota}\left(pr\cos\left(\theta-\phi\right)+\ell\phi\right)}.
    \end{split}
\end{align}
The angular integral can be simplified as follows:
\begin{equation}
    \int_{0}^{2\pi}d\phi\:\exp\left[\dot{\iota}\left(pr\cos\left(\phi-\theta\right)+\ell\phi\right)\right]=\exp\left[\dot{\iota}\ell\theta\right]\int_{0}^{2\pi}d\tilde{\phi}\:\exp\left[\dot{\iota}\left(pr\cos\left(\tilde{\phi}\right)+\ell\tilde{\phi}\right)\right].
\end{equation}
We use the following Jacobi–Anger expansion \cite{arfken1999mathematical} to perform the above angular integration:
\begin{equation}
\exp\left[\dot{\iota}a\cos x\right]=\sum_{n=-\infty}^{\infty}\dot{\iota}^{n}J_{n}\left(a\right)\exp\left[\dot{\iota}nx\right],
\end{equation}
where $J_{n}$ is the $n^{th}$ Bessel function of the first kind. The result of the angular integral is given by:
\begin{equation}
\int_{0}^{2\pi}d\tilde{\phi}\:\exp\left[\dot{\iota}\left(pr\cos\left(\tilde{\phi}\right)+\ell\tilde{\phi}\right)\right]=2\pi\dot{\iota}^{-\ell}\left(-1\right)^{\ell}J_{\ell}\left(pr\right).
\end{equation}
Ignoring the phase factors, we have:
\begin{equation}
 \tilde{\psi}_\ell\left(r,\theta\right)=\exp\left[\dot{\iota}\ell\theta\right]\left(\frac{2\sqrt{\pi}}{\sigma}\right)^{1/2}\frac{1}{2\pi}\int_{0}^{\infty}dp\sqrt{p}\exp\left[-\frac{\left(p-p_{0}\right)^{2}}{2\sigma^{2}}\right]J_{\ell}\left(pr\right).
\end{equation}
To perform this integral, we again introduce the dimensionless variables $x$ and $u_0$ ($u_0\equiv p_0/\sigma$ and $x\equiv (p-p_0)/\sigma$) that we used before:
\begin{equation}
\int_{0}^{\infty}dp\sqrt{p}\exp\left[-\frac{\left(p-p_{0}\right)^{2}}{2\sigma^{2}}\right]J_{\ell}\left(pr\right)=\sigma^{3/2}\int_{-u_{0}}^{\infty}dx\sqrt{u_{0}+x}\exp\left[-\frac{x^{2}}{2}\right]J_{\ell}\left(\left(u_{0}+x\right)\sigma r\right).
\end{equation}
For large $x$, $J_{\ell}\left(x\right)$ has the following asymptotic form:
\begin{equation}
J_{\ell}\left(x\right)\approx\sqrt{\frac{2}{\pi x}}\cos\left[x-\frac{\ell\pi}{2}-\frac{\pi}{4}\right].
\end{equation}
This approximation is justified as long as $p_{0}r\gg1$. Using the asymptotic expansion, we have:
\begin{equation}
\int_{-u_{0}}^{\infty}dx\sqrt{u_{0}+x}\exp\left[-\frac{x^{2}}{2}\right]J_{\ell}\left(\left(u_{0}+x\right)\sigma r\right)\approx\sqrt{\frac{2}{\pi\sigma r}}\cos\left[u_{0}\sigma r-\frac{\ell\pi}{2}-\frac{\pi}{4}\right]\int_{-\infty}^{\infty}dx\exp\left[-\frac{x^{2}}{2}\right]\cos\left[x\sigma r\right].
\end{equation}
Using
\begin{equation}
\int_{-\infty}^{\infty}dx\cos\left[ax\right]e^{-\frac{x^{2}}{2}}=\sqrt{2\pi}\exp\left[-\frac{a^{2}}{2}\right],
\end{equation}
we arrive at the final expression:
\begin{equation}
\tilde{\psi}_\ell\left(r,\theta\right) = \left(\frac{p_{0}\sigma}{\sqrt{\pi}}\right)^{1/2}J_{\ell}\left(p_{0}r\right)\exp\left[-\frac{\sigma^{2}r^{2}}{2}\right]\exp\left[\dot{\iota}\ell\theta\right],
\label{eq: real_psi_1}
\end{equation}
or equivalently
\begin{equation}
\tilde{\psi}_\ell\left(r,\theta\right)=\left(\frac{2\sigma}{\pi\sqrt{\pi}}\right)^{1/2}\frac{\cos\left[p_{0}r-\frac{\ell\pi}{2}-\frac{\pi}{4}\right]}{\sqrt{r}}\exp\left[-\frac{\sigma^{2}r^{2}}{2}\right]\exp\left[\dot{\iota}\ell\theta\right].
\label{eq: real_psi_2}
\end{equation}
\section{Harmonic Oscillator with Mexican hat dispersion and Berry curvature}
\label{sec:MHHOBC}

Here we derive the eigenenergies of an electron with a Mexican hat dispersion and Berry curvature that is subject to a harmonic confining potential.
To incorporate the Berry curvature, the Appendix \ref{sec:HO_BC} provides the recipe that in momentum space $\hat{\vec{r}}^2$ should be replaced by $\left(\dot{\iota}\vec{\nabla}_{\vec{q}}+A\left(\vec{q}\right)\right)^{2}$. Exploiting the rotational symmetry of the system, in the Coulomb gauge, the Berry connection can be written as
\begin{equation}
\vec{A}=\frac{\tilde{\Phi}\left(p\right)}{p}\hat{\phi},
\end{equation}
where
\begin{equation}
\tilde{\Phi}\left(p\right)=\int_{0}^{p}dp'\,p'\,\Omega\left(p'\right),
\end{equation}
is the amount of fractional Berry flux (the net Berry flux is given by $\Phi(p)=2\pi\tilde{\Phi}(p)$ ) through a disk in momentum space of radius $p$. Modifying the SE to incorporate the effect of Berry curvature amount to rewriting the azimuthal part of the gradient operator in the following way
\begin{equation}
\frac{\dot{\iota}}{p}\frac{\partial}{\partial\phi_{p}}\longrightarrow\frac{\dot{\iota}}{p}\frac{\partial}{\partial\phi_{p}}+\frac{\tilde{\Phi}\left(p\right)}{p}.
\end{equation}
The effective one-dimensional harmonic oscillator SE in this situation can be shown to be  (following the same procedure as in Appendix \ref{sec: MHHO})
\begin{equation}
\left[-\frac{d^{2}}{du^{2}}+\frac{1}{u^{2}}\left(-\frac{1}{4}+\left(-\ell+\tilde{\Phi}\left(u\sigma\right)\right)^{2}\right)+\left(u-u_{0}\right)^{2}\right]g=\epsilon g.
\label{eq:1DSE_BC}
\end{equation}
Let us again expand around $u_0$, changing the variable, $x\equiv u-u_{0}$
\begin{equation}
-\left[\frac{d^{2}g}{dx^{2}}+\frac{1}{u_{0}^{2}}\left(\frac{1}{4}-\left(-\ell+\tilde{\Phi}\left(p_{0}\right)\right)^{2}\right)g\right]+x^{2}g=\epsilon g.
\end{equation}
With the following definition
\begin{equation}
\tilde{\epsilon}_\ell=\epsilon+\frac{1}{4u_{0}^{2}}-\frac{\left(-l+a\right)^{2}}{u_{0}^{2}},
\end{equation}
we get the following familiar differential equation
\begin{equation}
-\frac{d^{2}g}{dx^{2}}+x^{2}g=\tilde{\epsilon}_{\ell}g.
\end{equation}
We once again encounter the 1D simple harmonic oscillator SE, wherein eigenvalues are given by
\be
\tilde{\epsilon}_{n,l}=\left(2n+1\right).
\ee
\section{Exchange Interaction between two Gaussian wavepackets with cosine}
\label{sec: exchange}

Here we provide a naive estimation of the exchange interaction between two neighboring electrons in the WC, using the wavefunction corresponding to the Mexican hat dispersion (Eq.~\ref{eq:rs_wf}). As emphasized in Sec.~\ref{sec: exchange_order}, this calculation provides only a lower bound for the exchange energy since higher-order ring exchange processes are typically dominant in the WC state \cite{Roger1984Multiple, chakravarty1999wigner, Katano2000Multiple, Kim2022interstitial}.

We assume that we have two wave packets given by Eq.~\ref{eq: real_psi_2} are located at $\vec{r} = \vec{0}$ and $\vec{r} = \vec{R}$. We now try to estimate the exchange interaction between those wavepackets by computing the following integral
\begin{equation}
J_{ab}=\int d\vec{r_{1}}\int d\vec{r_{2}}\;\left[\psi^{*}\left(\vec{r}_{1}\right)\psi^{*}\left(\vec{r}_{2}-\vec{R}\right)V\left(\vec{r}_{1}-\vec{r}_{2}\right)\psi\left(\vec{r}_{2}\right)\psi\left(\vec{r}_{1}-\vec{R}\right)\right].
\end{equation}
Substituting the wavefunctions, we get
\begin{align}
    \begin{split}
        J_{ab}=&\left(\frac{2\sigma}{\pi\sqrt{\pi}}\right)^{2}\int d\vec{r_{1}}\int d\vec{r_{2}}\left(\frac{\cos\left(p_{0}\left|\vec{r_{1}}\right|-\frac{\ell\pi}{2}-\frac{\pi}{4}\right)}{\sqrt{\left|\vec{r_{1}}\right|}}\frac{\cos\left(p_{0}\left|\vec{r_{2}}-\vec{R}\right|-\frac{\ell\pi}{2}-\frac{\pi}{4}\right)}{\sqrt{\left|\vec{r_{2}}-\vec{R}\right|}}\exp\left(-\frac{\sigma^{2}\vec{r}_{1}^{2}}{2}\right)\exp\left(-\frac{\sigma^{2}\left(\vec{r}_{2}-\vec{R}\right)^{2}}{2}\right)V\left(\vec{r}_{1}-\vec{r}_{2}\right)\right.\\&\left.\frac{\cos\left(p_{0}\left|\vec{r_{2}}\right|-\frac{\ell\pi}{2}-\frac{\pi}{4}\right)}{\sqrt{\left|\vec{r_{2}}\right|}}\frac{\cos\left(p_{0}\left|\vec{r_{1}}-\vec{R}\right|-\frac{\ell\pi}{2}-\frac{\pi}{4}\right)}{\sqrt{\left|\vec{r_{1}}-\vec{R}\right|}}\exp\left(-\frac{\sigma^{2}\vec{r}_{2}^{2}}{2}\right)\exp\left(-\frac{\sigma^{2}\left(\vec{r}_{1}-\vec{R}\right)^{2}}{2}\right)\right).
    \end{split}
\end{align}
The major contribution to the integrand comes when the Gaussian overlap is the maximum. Geometrically this corresponds to the midway between these two wavepackets. With this intuition in mind, let us define $\vec{r}_{1}=\frac{\vec{R}}{2}+\vec{\epsilon}_{1}$ and $\vec{r_{2}}=\frac{\vec{R}}{2}+\vec{\epsilon}_{2}$ so that
\begin{align}
    \begin{split}
        \exp\left(-\frac{\sigma^{2}\vec{r}_{1}^{2}}{2}\right)\exp\left(-\frac{\sigma^{2}\left(\vec{r}_{1}-\vec{R}\right)^{2}}{2}\right)&=\exp\left[-\frac{\sigma^{2}R^{2}}{4}\right]\exp\left[-\sigma^{2}\vec{\epsilon}_{1}^{2}\right],\\\exp\left(-\frac{\sigma^{2}\vec{r}_{2}^{2}}{2}\right)\exp\left(-\frac{\sigma^{2}\left(\vec{r}_{2}-\vec{R}\right)^{2}}{2}\right)&=\exp\left[-\frac{\sigma^{2}R^{2}}{4}\right]\exp\left[-\sigma^{2}\vec{\epsilon}_{2}^{2}\right].
    \end{split}
\end{align}
The exchange integral can be further simplified 
\begin{align}
    \begin{split}
        J_{ab}&=\left(\frac{2\sigma}{\pi\sqrt{\pi}}\right)^{2}\exp\left[-\frac{\sigma^{2}R^{2}}{2}\right]\int d\vec{\epsilon_{1}}\int d\vec{\epsilon_{2}}\left(\frac{\cos\left(p_{0}\left|\frac{\vec{R}}{2}+\vec{\epsilon}_{1}\right|-\frac{\ell\pi}{2}-\frac{\pi}{4}\right)}{\sqrt{\left|\frac{\vec{R}}{2}+\vec{\epsilon}_{1}\right|}}\frac{\cos\left(p_{0}\left|\frac{\vec{R}}{2}-\vec{\epsilon}_{2}\right|-\frac{\ell\pi}{2}-\frac{\pi}{4}\right)}{\sqrt{\left|\frac{\vec{R}}{2}-\vec{\epsilon}_{2}\right|}}\exp\left[-\sigma^{2}\vec{\epsilon}_{1}^{2}\right]V\left(\vec{\epsilon}_{1}-\vec{\epsilon}_{2}\right)\right.\\&\left.\frac{\cos\left(p_{0}\left|\frac{\vec{R}}{2}+\vec{\epsilon}_{2}\right|-\frac{\ell\pi}{2}-\frac{\pi}{4}\right)}{\sqrt{\left|\frac{\vec{R}}{2}+\vec{\epsilon}_{2}\right|}}\frac{\cos\left(p_{0}\left|\frac{\vec{R}}{2}-\vec{\epsilon}_{1}\right|-\frac{\ell\pi}{2}-\frac{\pi}{4}\right)}{\sqrt{\left|\frac{\vec{R}}{2}-\vec{\epsilon}_{1}\right|}}\exp\left[-\sigma^{2}\vec{\epsilon}_{2}^{2}\right]\right),\\&\approx\left(\frac{2\sigma}{\pi\sqrt{\pi}}\right)^{2}\exp\left[-\frac{\sigma^{2}R^{2}}{2}\right]\left(\frac{2}{R}\right)^{2}\int d\vec{\epsilon_{1}}\int d\vec{\epsilon_{2}}\left(\cos\left(p_{0}\left|\frac{\vec{R}}{2}+\vec{\epsilon}_{1}\right|-\frac{\ell\pi}{2}-\frac{\pi}{4}\right)\cos\left(p_{0}\left|\frac{\vec{R}}{2}-\vec{\epsilon}_{2}\right|-\frac{\ell\pi}{2}-\frac{\pi}{4}\right)\exp\left[-\sigma^{2}\vec{\epsilon}_{1}^{2}\right]\right.\\&\left.V\left(\vec{\epsilon}_{1}-\vec{\epsilon}_{2}\right)\cos\left(p_{0}\left|\frac{\vec{R}}{2}+\vec{\epsilon}_{2}\right|-\frac{\ell\pi}{2}-\frac{\pi}{4}\right)\cos\left(p_{0}\left|\frac{\vec{R}}{2}-\vec{\epsilon}_{1}\right|-\frac{\ell\pi}{2}-\frac{\pi}{4}\right)\exp\left[-\sigma^{2}\vec{\epsilon}_{2}^{2}\right]\right).
    \end{split}
\end{align}
We can use the following cosine identity to simplify the above integral
\begin{equation}
\cos A\cos B=\frac{1}{2}\left(\cos\left(A+B\right)+\cos\left(A-B\right)\right).
\end{equation}
Also treating $\left|\vec{\epsilon}_{1,2}\right|\ll\left|\vec{R}\right|$, we get $|\vec{R}/2\pm\vec{\epsilon}_{1,2}|\approx R/2\pm\vec{R}\cdot\vec{\epsilon}_{1,2}/R$. Thus, we arrive at the following expression
\begin{equation}
J_{ab}\approx\left(\frac{2\sigma}{\pi\sqrt{\pi}}\right)^{2}\exp\left[-\frac{\sigma^{2}R^{2}}{2}\right]\left(\frac{2}{R}\right)^{2}\left(\frac{\cos\left(p_{0}R-\ell\pi-\frac{\pi}{2}\right)}{2}\right)^{2}\int d\vec{\epsilon_{1}}\int d\vec{\epsilon_{2}}\exp\left[-\sigma^{2}\vec{\epsilon}_{1}^{2}\right]\exp\left[-\sigma^{2}\vec{\epsilon}_{2}^{2}\right]V\left(\vec{\epsilon}_{1}-\vec{\epsilon}_{2}\right).
\end{equation}
Under a change of variable $\vec{\alpha_1} = \vec{\epsilon_1} + \vec{\epsilon_2}$ and $\vec{\alpha_2} = \vec{\epsilon_1} - \vec{\epsilon_2}$ we can convert the above integral into
\begin{align}
    \begin{split}
        \int d\vec{\epsilon_{1}}\int d\vec{\epsilon_{2}}\exp\left[-\sigma^{2}\vec{\epsilon}_{1}^{2}\right]\exp\left[-\sigma^{2}\vec{\epsilon}_{2}^{2}\right]V\left(\vec{\epsilon}_{1}-\vec{\epsilon}_{2}\right)&=\int d\vec{\epsilon_{1}}\int d\vec{\epsilon_{2}}V\left(\vec{\epsilon}_{1}-\vec{\epsilon}_{2}\right)\exp\left[-\frac{\sigma^{2}\left(\vec{\epsilon_{1}}+\vec{\epsilon_{2}}\right)^{2}}{2}\right]\exp\left[-\frac{\sigma^{2}\left(\vec{\epsilon_{1}}-\vec{\epsilon_{2}}\right)^{2}}{2}\right],\\&=\int d\vec{\alpha_{1}}\left|\frac{\partial\left(\alpha_{1x},\alpha_{1y},\alpha_{2x},\alpha_{2y}\right)}{\partial\left(\epsilon_{1x},\epsilon_{1y},\epsilon_{2x},\epsilon_{2y}\right)}\right|^{-1}\exp\left[-\frac{\sigma^{2}\vec{\alpha}_{1}^{2}}{2}\right]\int d\vec{\alpha_{2}}V\left(\vec{\alpha_{2}}\right)\exp\left[-\frac{\sigma^{2}\vec{\alpha_{2}}^{2}}{2}\right],\\&=\frac{1}{4}\int d\vec{\alpha}_{1}\exp\left[-\frac{\sigma^{2}\vec{\alpha_{1}}^{2}}{2}\right]\int d\vec{\alpha_{2}}V\left(\vec{\alpha_{2}}\right)\exp\left[-\frac{\sigma^{2}\vec{\alpha_{2}}^{2}}{2}\right],\\&=\frac{1}{4}\left(\sqrt{\frac{2\pi}{\sigma^{2}}}\right)^{2}\int d\vec{\alpha_{2}}V\left(\vec{\alpha_{2}}\right)\exp\left[-\frac{\sigma^{2}\vec{\alpha_{2}}^{2}}{2}\right].
    \end{split}
\end{align}
Let us explicitly consider the Coulomb interaction, $V\left(r\right)= e^{2}/\epsilon_{r}r$ to evaluate the above integral.
\begin{align}
    \begin{split}
        \int d\vec{\alpha_2}\frac{1}{\alpha_2}\exp\left[-\frac{\sigma^{2}\vec{\alpha_2}^{2}}{2}\right]&=2\pi\int_{0}^{\infty}d\alpha_2\exp\left[-\frac{\sigma^{2}\alpha_2^{2}}{2}\right],\\&=\pi\sqrt{\frac{2\pi}{\sigma^{2}}}.
    \end{split}
\end{align}
Finally, arriving at the following result
\begin{equation}
J_{ab}\approx\sqrt{\frac{8}{\pi}}\left(\frac{1}{R\sigma}\right)^{2}\left(\frac{e^{2}\sigma}{\epsilon_{r}}\right)\left(\cos\left(p_{0}R-\ell\pi-\frac{\pi}{2}\right)\right)^{2}\exp\left[-\frac{\sigma^{2}R^{2}}{2}\right].
\label{eq: JabcosApp}
\end{equation}
For the sake of completeness, we  present the result for exchange interaction between two Gaussian wavepackets with similar widths and situated at the same position $\vec{r} = \vec{0}$ and $\vec{r} = \vec{R}$. The wavepackets are given by
\begin{equation}
\psi\left(\vec{r}\right)=\frac{1}{2\pi}\left(\frac{\sigma^{2}}{\pi}\right)^{\frac{1}{2}}\exp\left(-\frac{\vec{r}^{2}\sigma^{2}}{2}\right).
\end{equation}
Following a similar calculation as presented before, we can arrive at the final (exact) result
\begin{equation}
J_{ab} = \sqrt{\frac{\pi}{2}}\left(\frac{e^{2}\sigma}{\epsilon_{r}}\right)\exp\left[-\frac{R^{2}\sigma^{2}}{2}\right].
\label{eq: Jab}
\end{equation}
Comparing Eq.~\ref{eq: JabcosApp} and Eq.~\ref{eq: Jab} and recalling the definition of the Lindemann ratio $\eta \sim 1\Big/R\sigma$, we can make the following remarks.  
\begin{itemize}
    \item For a wavepacket with Mexican hat dispersion, $J_{ab}$ oscillates in magnitude (but is always nonnegative $J_{ab}$), leading to non-monotonic behavior in the ordering temperature as  a function of $n$.
    \item Ignoring the oscillatory $\cos^2$ term, $J_{ab}$ for the wavepacket with Mexican hat dispersion is smaller by a factor of $\eta^2$ compared to the usual Gaussian wave packets (i.e., to the case of a parabolic dispersion). This reduced $J_{ab}$ implies the exchange ordering is suppressed to a much lower temperature.
\end{itemize}

\section{Orbital Magnetization}
\label{sec: OrbMag}

\begin{figure}[htb]
\centering
\includegraphics[width=0.6 \columnwidth]{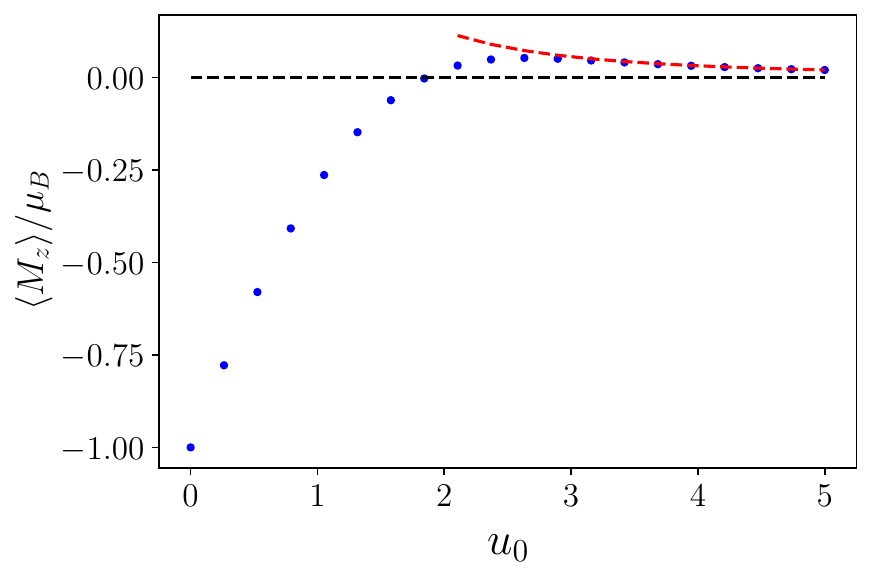}
\caption{Average magnetization $\langle\hat{M}_z\rangle$ of the first excited state ($\ell=1$) of an electron with Mexican hat dispersion as a function of the dimensionless Mexican hat radius $u_0$. Blue dots show results from a numeric calculation. The limit $u_0=0$ corresponds to the usual parabolic dispersion and gives the usual value $\langle\hat{M}_z\rangle=\mu_{B}$, resembling the case of a conventional 2D HO. As the radius increases, the magnetization decreases in magnitude and undergoes a sign change around $u_0 \simeq 1$. At larger values of $u_0$, $\langle\hat{M}_z\rangle$ converges towards the analytical prediction of Eq.~\ref{eq: mag_ana} (represented by a dashed red line).} 
\label{fig: Sign_change}
\end{figure}

Here we discuss the orbital magnetization of the WC state, focusing on the Mexican hat dispersion. We show how the sign of the magnetization is inverted with increasing $p_0$. Orbital magnetization is the negative of the partial derivative of free energy with respect to an applied magnetic field (which is a linear response to an external magnetic field).
\begin{equation}
\hat{M}_z=\left.-\frac{\partial F}{\partial B}\right|_{B=0}.
\end{equation}
At zero temperature and using minimal substitution, we can calculate the free energy as
\begin{align}
    \begin{split}
        F&=H\left(\vec{p}+e\vec{A}\right),\\&\simeq H\left(\vec{p}\right)+\frac{e}{2}\left(\vec{A}\cdot\vec{\nabla}_{\vec{p}}H+\left(\vec{\nabla}_{\vec{p}}H\right)\cdot\vec{A}\right).
    \end{split}
\end{align}
where $\vec{\nabla}_{\vec{p}}H=\vec{v}_{p}$ is the velocity operator and $-e$ is the charge of the particle. For a  constant magnetic field $\vec{B}=B\hat{z}$, and choosing $\vec{A}=-\frac{1}{2}\left(\vec{r}\times\vec{B}\right)$ we can derive
\begin{equation}
F\simeq H\left(\vec{p}\right)+\frac{e\vec{B}\cdot}{2}\left(\left(\frac{1}{2}\left(\vec{r}\times\vec{v}_{p}\right)\right)-\left(\frac{1}{2}\left(\vec{v}_{p}\times\vec{r}\right)\right)\right).
\end{equation}
Using the fact that $\left(\vec{r}\times\vec{v}_{p}\right)_{z}=-\left(\vec{v}_{p}\times\vec{r}\right)_{z}$, the magnetization operator can be found to be 
\begin{equation}
\hat{M}_z=\frac{e}{2}\left(\vec{v}_{p}\times\vec{r}\right)\hat{z}.
\end{equation}
For rotationally symmetric system $v_{\vec{p}}=v_{p}\hat{p}$, where $v_{p}=\frac{\partial E}{\partial p}$ and in the presence of Berry curvature, $m$ can be written 
\begin{equation}
\hat{M}_z = \frac{e}{2}\left(\frac{v_{p}}{p}\left(\dot{\iota}\frac{\partial}{\partial\phi}+\tilde{\Phi}\left(p\right)\right)\right).
\end{equation}
When $\psi\left(\vec{p}\right)=f\left(p\right)e^{\dot{\iota}\ell\phi}$, we get
\begin{equation}
\left\langle \hat{M}_z\right\rangle =\frac{e}{2}\,\frac{1}{\left(2\pi\right)^2}\int_{0}^{2\pi}d\phi\int_{0}^{\infty}dp\,v_{p}\left(-\ell+\tilde{\Phi}\left(p\right)\right)\left|f\left(p\right)\right|^{2}.
\end{equation}
Plugging the wavefunction from Eq.~\ref{eq:psi_p}, and expanding $\tilde{\Phi}\left(p\right)\simeq\tilde{\Phi}\left(p_{0}\right)+\left(p-p_{0}\right)p_{0}\Omega\left(p_{0}\right)$, one arrives at
\begin{equation}
\left\langle \hat{M}_z\right\rangle =\frac{e}{2m}\left(\frac{1}{\sqrt{\pi}\sigma}\right)\int_{0}^{\infty}dp\,\left(\frac{p-p_{0}}{p}\right)\exp\left[-\frac{\left(p-p_{0}\right)^{2}}{\sigma^{2}}\right]\left(-\ell+\tilde{\Phi}\left(p_{0}\right)+\left(p-p_{0}\right)p_{0}\Omega\left(p_{0}\right)\right).
\end{equation}
The two integrals in the above equation can be calculated to the leading order as follows
\begin{align}
    \begin{split}
    \int_{0}^{\infty}dp\,\left(\frac{p-p_{0}}{p}\right)\exp\left[-\frac{\left(p-p_{0}\right)^{2}}{\sigma^{2}}\right]& \simeq -\frac{\sqrt{\pi}}{2}\frac{\sigma}{u_{0}^{2}},\\
\int_{0}^{\infty}dp\,\frac{\left(p-p_{0}\right)^{2}}{p}\exp\left[-\frac{\left(p-p_{0}\right)^{2}}{\sigma^{2}}\right]&\simeq\frac{\sqrt{\pi}}{2}\frac{\sigma^{2}}{u_{0}}.    
    \end{split}
\end{align}
Thus we finally arrive at
\begin{equation}
\left\langle \hat{M}_z\right\rangle = \frac{e}{2m}\left(\frac{\sigma^{2}}{2p_{0}^{2}}\right)\left(\ell-\tilde{\Phi}\left(p_{0}\right)+\Omega\left(p_{0}\right)p_{0}^{2}\right).
\label{eq: mag_ana}
\end{equation}
Even in the absence of Berry curvature, the aforementioned findings indicate an intriguing transition as the radius of the Mexican hat, denoted as $u_0$, is varied. In the case of small $u_0$, the sign of the magnetization aligns with the physical notion of a negatively charged particle circulating counterclockwise around a loop. However, as $u_0$ increases, the magnitude of the magnetization diminishes and undergoes a sign reversal beyond a critical value. This phenomenon is depicted in Figure ~\ref{fig: Sign_change}. 

\section{Winding number transition with trigonal warping}
\label{sec: winding}
In this section, we show that at the level of the HO description, there is still a magnetization transition at large $U$ in the presence of trigonal warping. For $t\neq 0$ (see Eq.~\ref{eq:dispersionTW}), the dispersion can be approximated by a sum of three anisotropic parabolas above the band edge:
\begin{equation}
\varepsilon_{i}\left(\vec{p}\right)\approx\frac{\left(p_{x}-p_{x_0}^{i}\right)^{2}}{2m_{x}}+\frac{\left(p_{y}-p_{y_0}^{i}\right)^{2}}{2m_{y}}.
\end{equation}
where $i$ labels the three pockets, ${i = 1,2,3}$, and $\left(p_{x_0}^{i}, p_{y_0}^{i}\right)$ are the coordinates of those pockets, given by 
\be 
\left(p_{x_0}^{i}, p_{y_0}^{i}\right) = p_0\times \left\{ \begin{array}{cc}
(1,0), & i = 1\\
\left(-\frac{1}{2},\frac{\sqrt{3}}{2}\right), & i = 2\\
\left(-\frac{1}{2},-\frac{\sqrt{3}}{2}\right), & i = 3
\end{array}\right. .
\ee 
The local coordinates of the second and third pockets are rotated $2\pi/3$ and $4\pi/3$ with respect to the first pocket in order to maintain the $C_3$ symmetry.

The values of $p_0$ and $m_x$ are the same as the $p_0$ and $m$ from Eq.~\ref{eq:p0andm}; the value of $m_y$ is given below.  The effective mass $m_y$ is given by
\begin{equation}
m_y=\frac{1+ U^2}{9 t\left(2+ U^2\right)^\frac{1}{2}},
\label{eq:mxandmy}
\end{equation} 
which diverges as $t\longrightarrow0$, recovering the Mexican hat dispersion of the previous section. 

In the context of the WC phase, the wavefunction of a single electron in the Wigner lattice resides in a superposition of the three pockets in momentum space. Within each pocket, the wavefunction takes the form of an anisotropic Gaussian wave packet, with energy 
\be
E_{0}=\frac{1}{2}\sqrt{\frac{k}{m_{x}}}+\frac{1}{2}\sqrt{\frac{k}{m_{y}}}.
\ee 
As in the problem of a multi-well potential, however, the electron can generally lower its energy by residing in a symmetric superposition of the three pockets. The single-electron problem can be modeled as a three-site problem in momentum space with the following matrix Hamiltonian:
\begin{equation}
\hat{h} = \left[\begin{array}{ccc}
E_{0} & -\tau & -\tau\\
-\tau & E_{0} & -\tau\\
-\tau & -\tau & E_{0}
\end{array}\right],
\end{equation}
where $\tau$ represents the tunneling matrix element between the pockets, which exponentially decreases ($\log \tau \sim -p_0^2/\sigma^2$) as the distance between the pockets increases (increased $U$).  The eigenvalues of the matrix $\hat{h}$ are
\begin{equation}
    \tilde{E} = E_{0}-2\tau\cos\left(\frac{2\pi w}{3}\right),
    \label{eq: Ene_Side}
\end{equation}
where $w\in\left\{ 0,1,-1\right\}$ are the winding numbers. The corresponding eigenvectors $\varphi$ are given by
\begin{equation}
\varphi = 
\left[\begin{array}{c}
1\\
1\\
1
\end{array}\right] ,\left[\begin{array}{c}
1\\
\zeta\\
\zeta^{2}
\end{array}\right],\left[\begin{array}{c}
1\\
\zeta^{2}\\
\zeta
\end{array}\right],
\end{equation}
where $\zeta=\exp{\left[\dot{\iota}2\pi/3\right]}$ is one of the cube roots of unity. The degenerate excited states ($w=\pm1$) have a winding number of $\pm1$.

In the presence of the Berry flux, however, the tunneling term $\tau$ is modified to incorporate the Berry flux threading inside the three sites. This modification leads to a Peierls substitution in the momentum space Hamiltonian in the following way:
\begin{equation}
\hat{h} =     \left[\begin{array}{ccc}
E_{0} & -\tau\exp\left[\frac{\dot{\iota}2\pi \tilde{\Phi}}{3}\right] & -\tau\exp\left[-\frac{\dot{\iota}2\pi \tilde{\Phi}}{3}\right]\\
-\tau\exp\left[-\frac{\dot{\iota}2\pi \tilde{\Phi}}{3}\right] & E_{0} & -\tau\exp\left[\frac{\dot{\iota}2\pi \tilde{\Phi}}{3}\right]\\
-\tau\exp\left[\frac{\dot{\iota}2\pi \tilde{\Phi}}{3}\right] & -\tau\exp\left[-\frac{\dot{\iota}2\pi \tilde{\Phi}}{3}\right] & E_{0}
\end{array}\right],
\end{equation}
as in the previous section, $\tilde{\Phi}$ is the fraction of the $2 \pi$ Berry flux that lies within the area between the three pockets. The new energy eigenvalues are given by
\begin{equation}
    \tilde{E}=E_{0}-2\tau\cos\left[\frac{2\pi\left(-w+\tilde{\Phi}\right)}{3}\right],
\end{equation}
with the same eigenvectors. Notice first that increasing $\tilde{\Phi}$ from zero breaks the degeneracy between $w=\pm1$ states. When $\tilde{\Phi}=1/2$, the $w=0$ and $w=+1$ states become degenerate, and at $\tilde{\Phi} > 1/2$ the $w=1$ state is the ground state. This phenomenon is the momentum space analog of the problem of the triatomic molecule (three hydrogen atoms near the vertices of an equilateral triangle) with an external magnetic field piercing through the triangle. This problem exhibits a similar transition in the winding number (real space wavefunction) as a function of increasing magnetic flux through the molecule.

\begin{figure}[htb]
\centering
\includegraphics[width=0.81 \columnwidth]{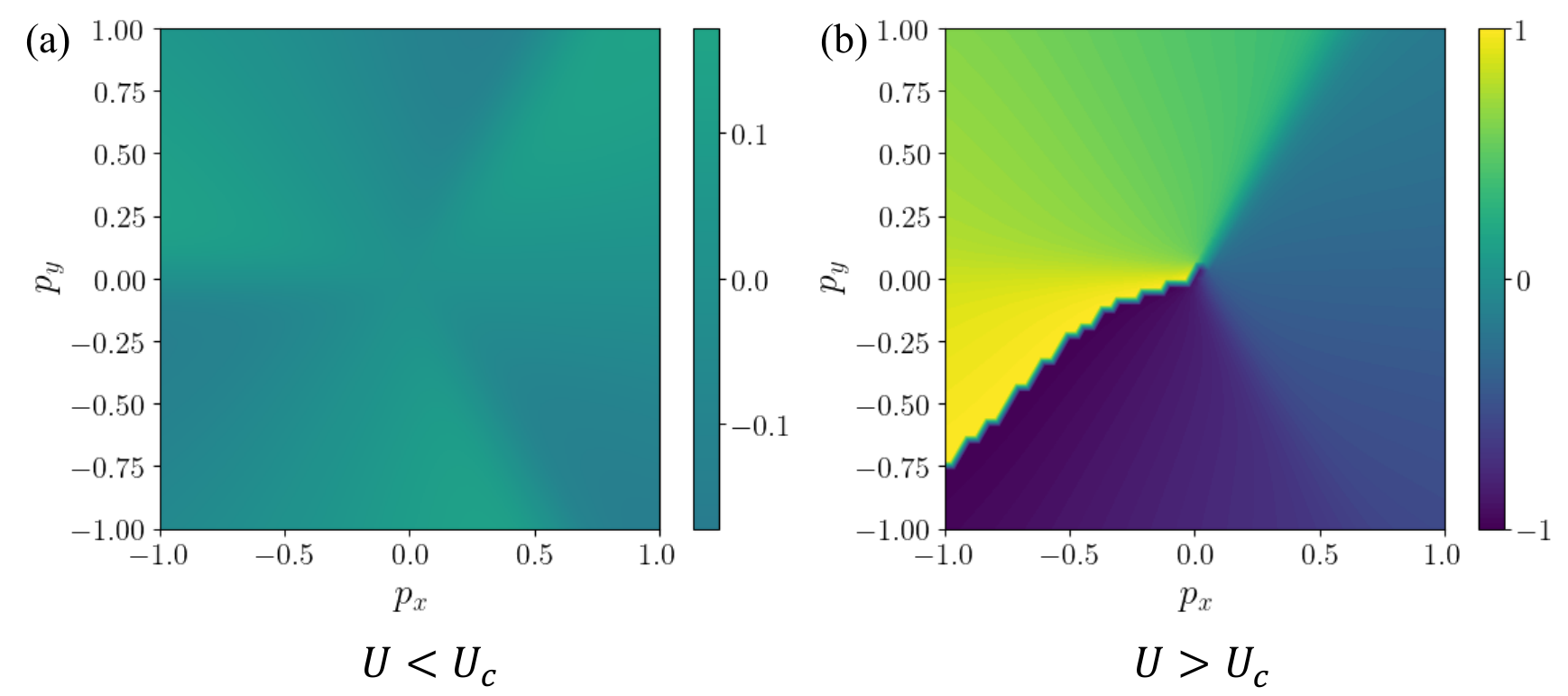}
\caption{Phase of the single particle wavefunction in the  momentum space is plotted for (a) $U<U_c$ and (b) $U>U_c$ for $n=10^{-2}$. The color bar is re-scaled from $(-\pi, \pi)$ to $(-1,1)$. The main text mentions that getting this winding transition for small $n$ where WC is stable is numerically challenging. Compared to $U<U_c$, when $U>U_c$, the phase has a jump of $2\pi$ corresponding to a phase winding.} 
\label{fig: Phase_Winding}
\end{figure}

Numerical verification of the winding number transition at large $U$ is difficult since at large $U$ the side pockets are well separated, and the effective tunneling $\tau$ between them is exponentially small. Hence, the splitting between different winding number states is also exponentially small. We can verify the transition numerically at a somewhat larger density $n$, which is technically beyond the WC phase's critical density but nonetheless indicates the presence of the winding transition. The results are shown in Fig.~\ref{fig: Phase_Winding}, which shows the change in the winding number of the wavefunction with increasing $U$.

\section{Numerical prescription for calculating Berry connection}
\label{sec: NumBConn}
To solve the Schrödinger equation in the presence of Berry curvature, one needs to be able to write down the operator $\left(\dot{\iota}\vec{\nabla}_{\vec{p}}+\vec{A}_m\left(\vec{p}\right)\right)^{2}$ in a discretized form, where the Berry connection $\vec{A}_{m}\left(\vec{q}\right)$ is given by
\begin{equation}
\vec{A}_{m}\left(\vec{q}\right)=\dot{\iota}\left\langle m,\vec{q}\right|\vec{\nabla}_{\vec{q}}\left|m,\vec{q}\right\rangle.
\end{equation}
However, numerically calculated eigenvectors $\left|m,\vec{q}\right\rangle$ may have arbitrary phase factors associated with them. To address this issue, we fix the phases of the eigenvectors such that $\langle m,\vec{q}|m,\vec{0}\rangle$ is a real number. This can be achieved by redefining each eigenvector as $\exp\left[\dot{\iota}\theta_{q_{x},q_{y}}\right]\left|m,\vec{q}\right\rangle$, where
\begin{equation}
\langle m,\vec{q}|m,\vec{0}\rangle=r_{q_{x},q_{y}}\exp\left[\dot{\iota}\theta_{q_{x},q_{y}}\right].
\end{equation}
With this redefined eigenvector, $\langle m,\vec{q}|m,\vec{0}\rangle$ becomes a real number. We use these redefined eigenvectors to numerically calculate the Berry connection $\vec{A}_{m}\left(\vec{q}\right)$. The BC calculated by first calculating the Berry connection this way is consistent with the BC results obtained from the gauge-invariant procedure.
\end{document}